\documentclass[iop,appendixfloats]{emulateapj}

\newcommand{\figurepath}{.}

\newbox\grsign \setbox\grsign=\hbox{$>$}
\newdimen\grdimen \grdimen=\ht\grsign
\newbox\laxbox \newbox\gaxbox
\setbox\gaxbox=\hbox{\raise.5ex\hbox{$>$}\llap
     {\lower.5ex\hbox{$\sim$}}}\ht1=\grdimen\dp1=0pt
\setbox\laxbox=\hbox{\raise.5ex\hbox{$<$}\llap
     {\lower.5ex\hbox{$\sim$}}}\ht2=\grdimen\dp2=0pt

\newcommand{\snr}{ \color{JungleGreen} }

\shorttitle{3D jet launching}
\shortauthors{Sheikhnezami \& Fendt}
\usepackage[usenames,dvipsnames]{color}
\usepackage{amsmath}
\usepackage{float}
\usepackage{ctable} 
\usepackage{color}
\begin{document}

\title{
Wobbling and precessing jets from warped disks in binary systems
}
\author{Somayeh Sheikhnezami
\altaffilmark{1,2} 
\&
        Christian Fendt
 \altaffilmark{2},
}
\altaffiltext{1}{School of Astronomy, Institute for Research in Fundamental Sciences (IPM), Tehran, Iran}
\altaffiltext{2}{Max Planck Institute for Astronomy, Heidelberg, Germany}
\email{snezami@ipm.ir, fendt@mpia.de}                                   

\begin{abstract}
We present results of the first ever three-dimensional (3D) magnetohydrodynamic (MHD)
simulations of the accretion-ejection structure.
We investigate the 3D evolution of jets launched symmetrically from single
stars but also jets from warped disks in binary systems.
We have applied various model setups and tested them by simulating a stable and bipolar symmetric 
3D structure from a single star-disk-jet system. 
Our reference simulation maintains a good axial symmetry and also a bipolar symmetry for more than 
600 rotations of the inner disk confirming the quality of our model setup.
We have then implemented a 3D gravitational potential (Roche potential) due by a companion star 
and run a variety of simulations with different binary separations and mass ratios.
These simulations show typical 3D deviations from axial symmetry, such as jet inclination outside the
Roche lobe or spiral arms forming in the accretion disk.
In order to find indication for precession effects, we have also run an exemplary parameter setup, 
essentially governed by a small binary separation of only $\simeq{200}$ inner disk radii. 
This simulation shows strong indication that we observe the onset of a jet precession caused by the
wobbling of the jet launching disk. 
We estimate the opening angle of the precession cone defined by the lateral motion of the the jet axis 
of about 4 degree after about 5000 dynamical time steps.

\end{abstract}
\keywords{
   accretion, accretion disks --
   MHD -- 
   ISM: jets and outflows --
   stars: pre-main sequence --
   galaxies: jets --
   galaxies: active 
 }
\section{Introduction}
Jets are powerful signatures of astrophysical activity and are observed over a wide range 
of luminosity and spatial scale.
Typical jet sources are young stellar objects (YSO), micro-quasars, and active galactic 
nuclei (AGN), while there is indication of jet motion also for a few pulsars, and for 
gamma-ray bursts
\citep{1974MNRAS.167P..31F, 1979Natur.279..701A, 1983ApJ...274L..83M, 
1994Natur.371...46M, 1997ApJ...487L...1R}.

Observations of jets from YSOs have revealed that the mass-loss carried by the jet is proportional 
to the disk accretion rate, suggesting a direct physical link between accretion and ejection
\citep{1990ApJ...354..687C, 1995ApJ...452..736H,
2006ApJ...646..319E, 2007IAUS..243..203C}.
Observational data also show the signatures of the magnetic field in the regions
where jets are formed \citep{1997Natur.385..415R, 2013EPJWC..6103003C},
as well as a considerable magnetization of the jet-launching object
(see e.g. \citealt{1990AJ.....99..946B, 2005ApJ...626..104M}).

It is now commonly accepted that magneto hydrodynamic (MHD) processes are essential
for the launching, acceleration and collimation of the outflows and jets from accretion disks
\citep{1982MNRAS.199..883B, 1983ApJ...274..677P, 1997A&A...319..340F, 2007prpl.conf..277P, 2015SSRv..tmp...53H}.
With {\em "launching"} we denote the actual transition from {\em accretion to ejection}, while
{\em "formation"} denotes the {\em acceleration and collimation} of a disk wind into a jet beam. 

Most early MHD simulations did concentrate on the jet formation problem.
In this case the disk evolution is not considered in the numerical treatment and the jet is
formed from a slow disk wind injected from the disk surface \citep{1995ApJ...439L..39U, 1997ApJ...482..712O}.
This approach is numerically less expensive and allows for a range of parameter studies. 
Furthermore, a number of physical processes could be included in the treatment and studied
concerning their impact on jet acceleration and collimation.
Examples are the role of 
radiative forces \citep{2011ApJ...742...56V}, 
magnetic diffusivity \citep{2002A&A...395.1045F, 2004IAUS..219..945C} and
relativity \citep{2009ApJ...705.1594M, 2010ApJ...709.1100P, 2011ApJ...737...42P}.
Further studies considered the radiative signatures of collimating MHD jets,
for example radiation maps for the forbidden emission lines \citep{2014A&A...562A.117T}
or polarized synchrotron radiation transfer \citep{2011ApJ...737...42P} in case of 
relativistic jets.
Also the impact of the disk magnetic field distribution and strength
\citep{2006ApJ...651..272F, 2006MNRAS.365.1131P, 2011ApJ...737...43F}, 
a magnetic jet confinement \citep{1986ApJ...311L..63C}, has been studied,
as well as 3D effects on jet formation \citep{2003ApJ...582..292O}.
On larger scales, interest has been gained recently (again) on the jet propagation and its feedback to the 
ISM or IGM (see e.g. \citealt{2011MNRAS.411..155G, 2014MNRAS.439.2903C}).

In order to understand what kind of disks launch jets and what kind of disks do not,
it is essential to include the disk physics in the treatment.
The first works on this subject followed an analytical approach based on a self-similar approach
\citep{1983ApJ...274..677P, 1985PASJ...37..515U, 1993ApJ...410..218W, 1995ApJ...444..848L, 1997A&A...319..340F}.
Today, numerical simulations of the accretion-ejection process play an essential role for the
understanding of jet launching. 
However, we note that it was already 1985 when the first jet launching simulations were published
\citep{1985PASJ...37..515U, 1986PASJ...38..631S}.
The accretion process in magnetized disks has been studied by numerical simulations first by \citet{1994ApJ...433..746S}.
Further pioneering work was presented by \citet{1998ApJ...508..186K} who were first in performing
simulations of jet launching from a {\em diffusive} MHD disk.
\citet{2002ApJ...581..988C, 2004ApJ...601...90C} and \citet{2007A&A...469..811Z}
extended such studies to considerably longer time scales and also to larger spatial 
scales, enabling them to derive the corresponding mass fluxes in disk and jet.

Following this - meanwhile - standard approach, further physical effects were investigated, such as 
the influence of the disk magnetization \citep{2009MNRAS.400..820T},
the launching from viscous disks \citep{2010A&A...512A..82M},
thermal effects \citep{2013MNRAS.428.3151T}, or even 
the launching of outflows from a magnetic field, self-generated by a mean-field disk
dynamo \citep{2004A&A...420...17V, 2003A&A...398..825V, 2014ApJ...796...29S}.

In all the launching simulations cited above, an axisymmetric setup was applied.
Concerning the launching and acceleration process alone, such a limitation is probably sufficient.
However, jets are not smooth but structured and many of them show a deviation from straight motion.
Seemingly helical trajectories have been observed, and typically an asymmetry between jet and
counter jet\footnote{ While in the literature the dominating part of the bipolar outflow is usually called the ``jet'', in our paper we will denote with ``jet'' the outflow in positive $z$ direction, and the outflow in negative $z$ direction with ``counter jet''.}.
For jet and counter jet an S-shape or C-shape large-scale alignment has been observed (see \citealt{Fendt1998}).

Furthermore, we know that stars may form as binaries (see section below). 
In close binary pairs the axial symmetry of the jet source may be disturbed substantially.
Bipolar jets forming in a binary system may be affected substantially by tidal forces and torques,
that might be visible as 3D effects in the jet structure and jet propagation.
Theoretical arguments suggest that astrophysical disks are warped whenever a misalignment is present 
in the system, or when a flat disk becomes unstable due to external forces\citep{2013MNRAS.433.2403O}.
External forces may generate density waves in the disk and vertical bending waves \citep{2013MNRAS.430..699R}.
Naturally, in a binary system, we may think that the disk around one of the stars is misaligned with 
respect to the orbital plane and, thus, is also subject to disk warping
\citep{1995MNRAS.274..987P, 2013MNRAS.433.2142F}.
Clearly, all these perturbations in the disk structure will potentially affect the jet launching.

In order to study those non-axisymmetric structure in jet-disk environments, 3D simulations of jet launching 
are essential.
Three-dimensional simulations have been applied to study either the evolution of the disk
\citep{2003ApJ...595.1009R,2011ApJ...735..122F, 2013EPJWC..4603004L,2013MNRAS.430..699R, 2014ASPC..488..141W},
or the jet
\citep{2003ApJ...582..292O, 2009MNRAS.399.1802R, 2010MNRAS.402....7M, 2013MNRAS.429.2482P, 2014MNRAS.439.2903C,  2011MNRAS.411..155G}.
However, a 3D simulation of the accretion-ejection structure has not yet been published\footnote{
The work of \citet{2014ApJ...784..121S} investigates the global 3D structure of accretion disk threaded by
a vertical magnetic field, but the simulation box in $\theta$-direction was too small to follow jet launching.}.

In our recent papers we investigated the axisymmetric launching process of an outflow from a magnetically 
diffusive accretion disk \citep{2012ApJ...757...65S, 2013ApJ...774...12F, 2014ApJ...793...31S}, 
including the evolution of a disk-dynamo that generates the jet-launching magnetic field \citep{2014ApJ...796...29S}.
In the present paper we extend the previous studies to three dimensions.
Our goals are,
\begin{itemize}
\item[1)] to develop a proper model setup for jet launching in 3D,
\item[2)] to study the stability and symmetry of the jet and counter jet in the initial launching area, and
\item[3)] to investigate 3D effects of jets launched in a binary system, such as jet inclination, 
          disk warping, or precession.
\end{itemize}

The paper is organized as follows; 
Section 2 summarizes a few observations indicating 3D effects in jet launching sources. 
Section 3 describes the model setup, in particular the 3D initial and boundary conditions.
Section 4 introduces to the 3D gravitational potential for the binary star-disk-jet system.
Results of 3D symmetric test runs are presented in section 5.
In section 6, we present and discuss the 3D simulations of jet launched from a binary system.

\section{Observations of jets in binary systems}
It is now well accepted that many stars form as binary or higher order multiple systems
\citep{2007ARA&A..45..565M}.
Confirmed binary systems that are sources of jets are T\,Tau 
\citep{1997A&AS..126..437H, 2002ApJ...568..771D, 2003AJ....125..858J} 
or RW\,Aur
\citep{1996AJ....111.2403H, 2012ARep...56..686B}.
Another source is HK\,Tau,
a binary system younger than four million years \citep{2014Natur.511..567J}.
Both stars, HK\,Tau\,B and HK\,Tau\,A, have circum-stellar disks that are misaligned with respect 
to the orbital plane of the binary.
Non-axisymmetric jet motion has been recently observed in HK\,Tau \citep{2014Natur.511..567J},
suggesting that one or both of the stellar disks may be inclined to the orbital 
plane \citep{2000MNRAS.317..773B}.

A further example is the spectroscopically identified bipolar jet of the pre-main sequence binary 
KH\,15D \citet{2010ApJ...708L...5M} that seems to be launched from the innermost part of the 
circum-binary disk, or may, alternatively, result from merging two outflows driven by the individual
stars, respectively.
This system is now known to be a young ($\sim 3$ Myr) eccentric binary system embedded in a nearly
edge-on circum-binary disk that is tilted, warped, and is precessing with respect to the binary orbit
\citep{2004AJ....127.2344J, 2004ApJ...607..913C, 2004AJ....128.1265J, 2004ApJ...603L..45W}.
The existence of a circum-binary disk in KH\,15D is evident from dust settling \citep{2010ApJ...711.1297L}.
An exceptionally striking example of a precessing jet from a X-ray binary is the classic source SS\,433
\citep{1979ApJ...233L..63M, 2015A&A...574A.143M}.
Arguments have been raised that in these systems it is the lack of axisymmetry that does not 
allow to produce a jet beam that is stable over a substantial propagation distance \citep{Fendt1998}.

Non-axisymmetric effects have also been found for an extra-galactic jet source.
Water maser observations of NGC\,4258 have detected a Keplerian rotation of the inner accretion 
disk \citep{1996ApJ...468L..17H, 2013MNRAS.436.1278W}.
Model fits of the disk kinematics clearly indicate disk warping on the scale of $8.6\times 10^4$ 
gravitational radii, corresponding to a size of the warped disk of about $\sim 0.04 {\rm pc}$
surrounding a super massive black hole of $\sim 3.9 \times 10^7 M_\odot$.
The inner radio is launched perpendicular to that disk \citep{1997ApJ...475L..17H}, although on 
large spatial scales three helical braids of jets can be disentangled spectroscopically \citep{1992ApJ...390..365C}. 
The exact interpretation of the large-scale kinematics is still unclear \citep{2007A&A...467.1037K}.

\section{Model approach}
For our numerical simulations, we apply the MHD code PLUTO version 4 
\citep{2007ApJS..170..228M, 2012ApJS..198....7M}
solving the conservative, time-dependent, resistive, inviscous MHD equations, namely for the
conservation of mass, momentum, and energy,
\begin{equation}
\frac{\partial\rho}{\partial t} + \nabla \cdot(\rho \vec v)=0,
\end{equation}
\begin{equation}
\frac{\partial(\rho \vec v)}{\partial t} + 
\nabla \cdot \left(\vec v \rho \vec v - \frac{\vec B \vec B}{4\pi} \right) + \nabla \left( P + \frac{B^2}{8\pi} \right)
+ \rho \nabla \Phi = 0,
\end{equation}
\begin{multline}
 \frac{\partial e}{\partial t} + \nabla \cdot \left[ \left( e + P + \frac{B^2}{8\pi} \right)\vec v 
 - \left(\vec v \cdot \vec B \right)\frac{\vec B}{4\pi} + ({\eta} \vec j ) \times \frac{\vec B}{4\pi}  \right]\\
 = - \Lambda_{\rm cool}. 
\end{multline}

Here, $\rho$ is the mass density, $\vec v$ the velocity, $P$ the thermal gas pressure,
$\vec B$ the magnetic field,
and $\Phi$ the 3D gravitational potential of the binary system (see section \ref{sec:3dGR}).
The electric current density $\vec j$ is given by Amp\'ere's law 
$\vec j = (\nabla \times \vec B) / 4\pi$.
The magnetic diffusivity can be  most generally defined as a tensor ${\bar{\eta}}$
(see our discussion in \citealt{2012ApJ...757...65S}).
In this paper, for simplicity we assume a scalar, isotropic magnetic diffusivity 
$\eta_{ij} \equiv \eta(x,y,z)$.
The evolution of the magnetic field is described by the induction equation,
\begin{equation}
\frac{\partial \vec B}{\partial t} - \nabla\times (\vec v \times \vec B - \eta \vec j) = 0.
\end{equation}

The cooling term $\Lambda$ in the energy equation can be expressed in terms of Ohmic heating
$\Lambda = g \Gamma$, with $\Gamma = ({\eta} \vec j) \cdot \vec j$, and with $g$ measuring the fraction of the 
magnetic energy that is radiated away instead of being dissipated locally. 
For simplicity, here we adopt again $g=1$. 
The gas pressure follows an equation of state $P = (\gamma - 1) u$ with the polytropic index 
$\gamma$ and the internal energy density $u$.
The total energy density is
\begin{equation}
e = \frac{P}{\gamma - 1} + \frac{\rho v^2}{2} + \frac{B^2}{2} + \rho \Phi.
\end{equation}

\subsection{Numerical setup}
Compared to the axisymmetric setup that has been applied to most launching simulations so far, 
the case of 3D simulations is substantially more demanding.
This holds for the technical treatment of the numeric as well as for the computational
resources.
Modifications have to be made for the initial conditions and the boundary conditions,
in particular when considering an orbiting binary system, or angular momentum 
conservation with in a rectangular grid.

The three-dimensional treatment of jet launching can be approached by two steps of complexity. 
The first step is to exploit the 3D-evolution of a jet launched from an axisymmetric setup,
in particular applying an axisymmetric gravitational potential.
The second step of complexity is to apply a non-axisymmetric setup on a priori, 
e.g. a non-axisymmetric gravitational potential of a binary system.
In this paper, we will apply both model setups, 
while we use step one primarily as a test case of our 3D setup.

\begin{figure}
\centering
\includegraphics[width=1\columnwidth]{\figurepath/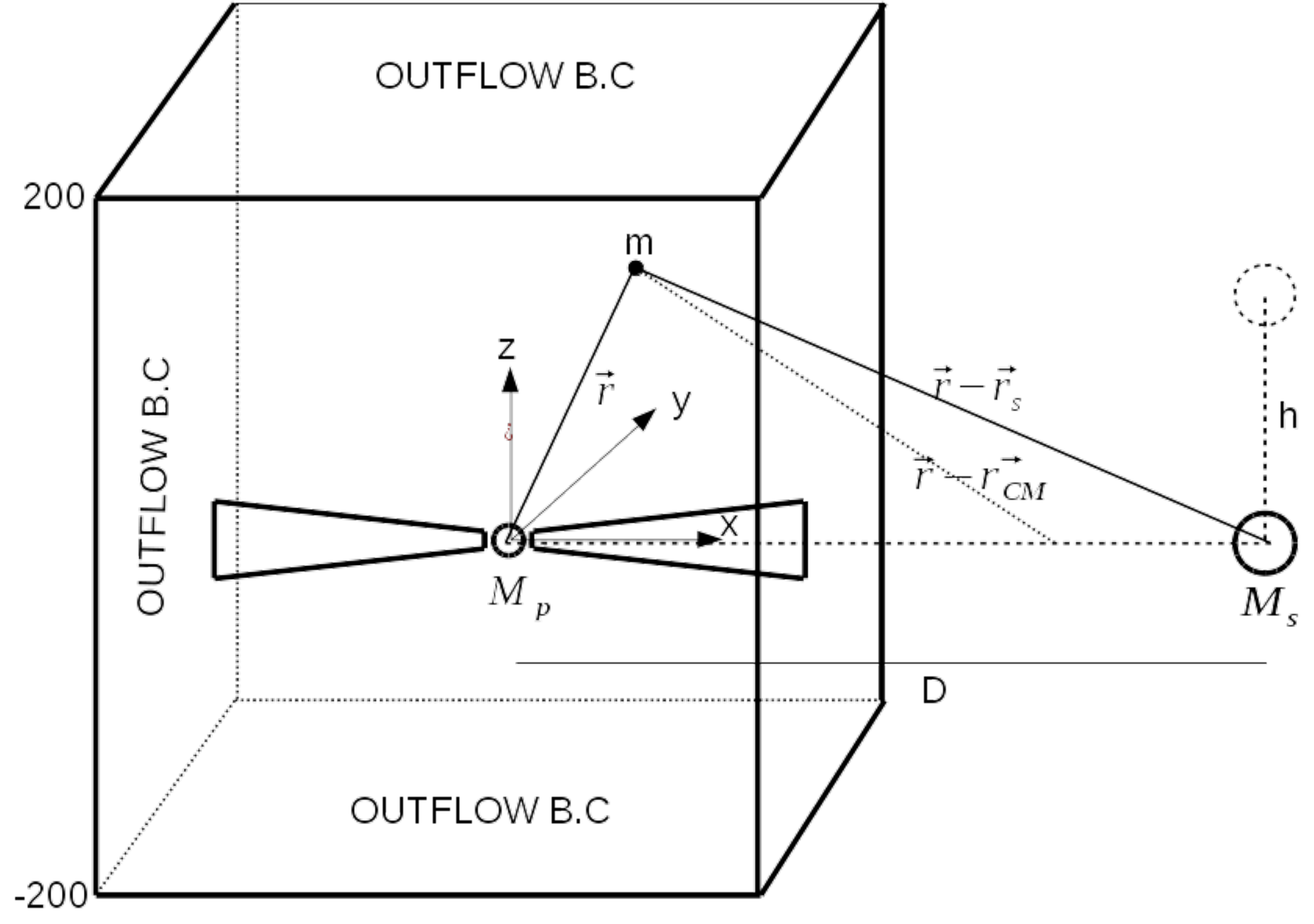}
\caption{Sketch of our model setup. 
A binary system consisting of a primary of mass $M_{\rm p}$ that is surrounded by an accretion 
disk and a secondary of mass $M_{\rm s}$ orbits around the center of mass (CM) located at
$r_{\rm CM}$.
An arbitrary mass element, $m$, located at $r$ is affected by the combined gravito-centrifugal force
of the Roche potential.
The origin of the coordinates system is in the center of the primary star.
The secondary is located outside the computational domain with a separation $D$
from the primary and offset from the initial disk-primary plane (the $x$-$y$ plane) by a 
distance $h$, implying an inclination between the orbital plane and the disk plane of an 
angle $\alpha = \arctan(h/D)$.}
\label{fig:binary_structure}
\end{figure}

Figure \ref{fig:binary_structure} illustrates the general setup for the simulations of a 
binary system.
With $M_{\rm p}$ and $M_{\rm s}$ we denote the mass of the primary and the secondary 
star. 
In our notation, it is the {\em primary star that is surrounded by a jet-launching accretion disk}.

The origin of the coordinate system is placed in the center of the primary star.
Both stars are orbiting the center of mass located at $r_{\rm CM}$ from the origin.
The horizontal and the vertical separation of two stars in that coordinates are denoted by 
the parameters $D$ and $h$, respectively.
The vertical separation implies an inclination between the orbital plane and the disk
plane with inclination angle $\alpha = \arctan(h/D)$. 

\begin{table} 
\caption {Characteristic parameters of our simulation runs. 
We list
the  mass ratio of the secondary to primary $q$,
the location of the secondary with respect to the primary, i.e. 
the vertical separation $h$ from the disk mid-plane and the binary separation $D$ along the disk midplane,
both resulting in an inclination between the orbital plane and the disk plane $\alpha$, 
the plasma beta at the inner disk radius $\beta$,
and the distance of the inner Lagrange point to the primary $L_1^{\rm p}$.
All scales are given in code units.
The magnetic diffusivity is parametrized by a coefficient indicating the magnitude of
diffusivity and the spatial profile $h_2$ or $h_3$, respectively.
Simulation {\em bcase2} applies an extremely small binary separation in order 
to be able to trace tidal effects in short simulation time scales.
}
\begin{center}
\begin{tabular}{ccccccc}
\hline
\hline
\noalign{\smallskip}
Run & $\eta$ & $q$ & $h$  & $D$ &  $\beta$ &  $L_1^{\rm p}\,[x,y,z]$\\
\noalign{\smallskip}
\hline
\noalign{\smallskip}
\multicolumn{7}{l}{Single star with jet launching disk }\\
\noalign{\smallskip}
\hline
\noalign{\smallskip}
 scase1 &   $0.03$            & - & -  & -    & 20 & -     \\
 scase2 &   $0.03$ for $z<10$ & - & -  & -    & 20 & -     \\
 scase3 &   $3h_2V_A$         & - & -  & -    & 20 & -     \\
 scase4 &   $0.03h_3$         & - & -  & -    & 20 & -    \\
 scase5 &   $0.03h_2$         & - & -  & -    & 20 & -    \\
 scase6 &   $0.03$ for $ z<5$ & - & -  & -    & 20 & -     \\
\noalign{\smallskip}
\hline
\hline
\noalign{\smallskip}  
\multicolumn{7}{l}{Binary system with jet launching disk around primary}\\
\noalign{\smallskip} 
\hline
\noalign{\smallskip}
bcase1  &  $0.03$ for $z<10$ & 2 & 60 & 300  & 20  & (130,0,26) \\
bcase2  &  $0.03$            & 1 & 60 & 200  & 20  & (100,0,30) \\
bcase3  &  $0.01$            & 2 & 60 & 300  & 20  & (130,0,26) \\
bcase4  &  $0.03$ for $z<10$ & 2 & 60 & 200  & 20  & (130,0,26) \\
\noalign{\smallskip}
\hline
 \end{tabular}
 \end{center}
\label{tbl:3D runs}
\end{table}

The parameters of the various simulation runs are shown in Table \ref{tbl:3D runs}.

We apply Cartesian coordinates $(x,y,z)$, and, contrary to axisymmetric simulations, 
the $z$-axis is not a symmetry axis any more.
Cartesian coordinates may cause problems when treating rotating objects (see discussions below),
however, they avoid artificial effects that may impose a symmetry of the system by boundary conditions
along the rotational axis (as for cylindrical or spherical coordinate systems).
Spherical coordinates are well suited for 3D disk simulations, as e.g. in \citet{2014ApJ...784..121S}, 
however, when investigating the 3D structure of a jet, a proper 3D treatment along the axis is  
essential.

The computational domain spans a cuboid with
the $z$ axis chosen along the direction of jet propagation.
The accretion disk mid-plane initially follows the $xy$-plane for $z=0$.
The computational domain typically extends over $ x\in [-120,120]$, $y\in [-120,120]$
and $z \in [-200,200]$ in units of the inner disk radius $r_{\rm i}$.

The numerical grid needs to be optimized in order to allow the best resolution for the 
physically most interesting parts of the computational domain.
We apply a uniform grid of $200 \times 200 \times 200 = 8\times 10^6$ cells for the
very inner part of the domain, $-5.0 < x,y,z < 5.0$.
For the rest part of the domain, a stretched grid is applied.
A total number of $400\times400\times560 = 8.96 \times 10^7$ grid cells are typically used 
for the whole computational domain,
although, we have also applied different physical sizes and grid resolutions for tests.

We apply the same units and normalization as in our previous papers \citep{2012ApJ...757...65S}.
Distances are expressed in units of the inner disk radius $r_{\rm i}$,
while $p_{\rm d,i}$ and $\rho_{\rm d,i}$ are the disk pressure and density at this radius, 
respectively\footnote{The index ''i'' refers to the value at the inner disk radius 
at the equatorial plane at time $t=0$}.

Velocities are normalized in units of the Keplerian velocity $v_{\rm K,i}$ at the inner disk 
radius.
We adopt $v_{\rm  K,i} = 1$ and $\rho_{\rm d,i} = 1$ in code units.
Time is measured in units of $t_{\rm i} = r_{\rm i} / v_{\rm K,i}$, which can be related to 
the Keplerian orbital period, $\tau_{\rm K,i} = 2\pi t_{\rm i}$.
The magnetic field is measured in units of $B_{\rm i} = B_{z,\rm i}$.

As usual we define the aspect ratio of the disk, $\epsilon$, as the ratio of the isothermal 
sound speed to the Keplerian speed, both evaluated at disk mid-plane, 
$\epsilon  \equiv {c_{\rm s}}/{v_{\rm K}}$.
Pressure is given in units of $p_{\rm d,i} = \epsilon^2 \rho_{\rm d,i} v_{\rm K,i}^2$.
Here $p_{\rm d,i} = \epsilon^2$ and $B_{\rm i} =\epsilon \sqrt{2/\beta }$ with the plasma-parameter 
$\beta$ is the ratio of thermal to magnetic pressure evaluated at the disk 
mid-plane\footnote{In PLUTO the magnetic field is normalized considering $4\pi = 1$}.

We apply the method of constrained transport (FCT) for the magnetic field evolution conserving
$\nabla \cdot B$ by definition.
For the spatial integration we use a linear algorithm with a second-order interpolation scheme,
together with the third-order Runge–Kutta scheme for the time evolution. 
Further, the HLL Riemann solver is chosen in our simulation.

\subsection{Initial state}
We generalize the initial axisymmetric setup used in our previous papers 
\citep{2012ApJ...757...65S, 2013ApJ...774...12F} into three dimensions.

We prescribe an initially geometrically thin disk with the thermal scale height $H$ and
$\epsilon = H/r =  0.1$. 
The accretion disk is in vertical equilibrium between the thermal 
pressure and the gravity (\citealt{1993A&A...276..625F, 2002ApJ...581..988C}).
A non-rotating corona is defined in pressure equilibrium with the disk.
The initial disk density distribution is

\begin{equation}
 \rho_{\rm d} = \rho_{\rm d,i}\left(\frac{2}{5\epsilon^2}\left[\frac{r_i}{R}-
          \left(1-\frac{5\epsilon^2}{2}\right) \frac{r_i}{r}\right] \right)^{3/2},
\end{equation}

while for the initial disk pressure distribution we apply
\begin{equation}
P_{\rm d} = P_{\rm d,i} \left(\frac{\rho_{\rm d,i}}{\rho_{\rm d}}\right)^{5/3}.
\end{equation}

Here, $r=\sqrt{x^2+y^2}$ and $R=\sqrt{x^2+y^2+z^2}$ denote the cylindrical and the spherical radius, respectively.
The accretion disk is set into a slightly sub-Keplerian rotation accounting for the radial gas pressure gradient 
and advection.

A deviation from Keplerian rotation can in principle also be due to field Lorentz forces.
Our initial field structure is not force-free, but since the plasma beta is rather high we can
neglect this effect.
Simulations of this and of our previous papers show that the disk-jet system will anyway 
establish a new {\em dynamic} equilibrium. 
This is a smooth process lasting about 100 inner disk rotations. 
The onset of an outflow does change the disk structure substantially, and the disk 
equilibrium will deviate from the initial distribution.

The initial magnetic field distribution is prescribed by the magnetic flux function $\psi$,
\begin{equation}
\psi(x,y,z)=3/4 B_{\rm z,i}r_{\rm i}^2(\frac{r}{r_{\rm i}})^{3/4}\frac{m^{5/4}}{{(m^2+(z/r)^2)}^{5/8}},
\end{equation}
where the parameter $m$ determines the magnetic field bending \citep{2007A&A...469..811Z}
 and in our model setup is set to the value, 0.4.
Here the $B_{\rm z,i}$ indicates to the vertical magnetic field at $(r=r_{\rm i},z=0)$.
Numerically, the poloidal field components are implemented by using the magnetic vector potential 
$A_{\phi}(x,y) = \psi/r$.
Initially $B_{\phi}=0$. 

We prescribe a non-rotating corona surrounding the disk.
This is in particular interesting in the case when we apply a finite disk radius $r_{\rm max}$,
implying that the accretion disk is embedded in an initially non-rotating corona.
This strategy was used by other authors before \citep{2001A&A...370..635B}.
The advantage of this method is that due to the vanishing rotation for large radii, 
no specific treatment is required at the outer grid boundary.
The disadvantage is that the mass reservoir for accretion is limited by the finite disk
mass. 
This may constrain the running time of the simulation as soon as the disk has lost a 
substantial fraction of its initial mass. 

However, since it is essential to treat the accretion process, properly,
we cannot use a similar strategy for the inner boundary and just neglect rotation, there.
Instead, a consistent rotational velocity must be assigned to the matter
at the inner boundary (see section \ref{sec:BC}).
The rotational velocity profile of the accretion disk is given by
\begin{equation}
  v_\phi (r) = \sqrt{\frac{GM}{r}} 
\begin{cases}
     0,  & {\rm for}\,\, 0 < r < r_0 \\
    \sqrt{1-5\epsilon^2}, & {\rm for} \,\, r_0 < r < r_{\rm i}\\
     \sqrt{1- 2.5\epsilon^2}, &{\rm for} \,\,  r_{\rm i} < r <  r_{\rm max}\\
     0, & {\rm for} \,\,r > r_{\rm max}
\end{cases}
\label{vphi_profile}
\end{equation}
where $r_{\rm i}$ denotes the inner disk radius and $r_0$ the inner radius of the ghost area corresponding 
to the inner boundary condition.
The radius $r_{\rm max}$ denotes the outer disk radius.

Above and below the disk, we define a density and pressure stratification that is in hydrostatic equilibrium with 
the gravity of the primary, a so-called {"}corona{"},
\begin{equation}
\rho_{\rm c}=\rho_{\rm a,i}\left(\frac{r_{\rm i}}{R}\right)^{1/(\gamma-1)}\!,\,
P_{\rm c}=\rho_{\rm a,i}\frac{\gamma-1}{\gamma}\frac{GM}{r_{\rm i}}\left(\frac{r_{\rm i}}{R}\right)^{\gamma/(\gamma-1)}\!.
\end{equation}
The parameter $\delta \equiv \rho_{\rm a,i} /\rho_{\rm d,i}$ quantifies the initial density contrast between 
disk and corona. In this paper $\delta = 10^{-4}$.

We note that in case of a binary system, the effective gravitational potential is given by the 3D Roche 
potential (see Section \ref{sec:3dGR}).
In such case, the outer parts of an initial corona as described above are not in hydrostatic equilibrium
anymore, as being affected by the gravity of the companion star and by centrifugal forces due to the orbital 
motion.
However, we find that we can safely neglect the 3D potential for the initial condition.
The initial corona will be swept out of the grid rather quickly and the new {\em dynamical} equilibrium for disk 
and outflow is governed by the 3D Roche potential.

\subsection{Boundary conditions}

The inner boundary plays an essential role for the evolution of the system.
In practice, it {``}hides{''} the gravitational singularity,
and absorbs the material that is delivered by the accretion disk.
We make use of the {\em internal boundary} option of PLUTO, that allows to prescribe
a structure of ghost cells within the active domain, that are updated by user defined
boundary values - in our case these boundary values allow to absorb disk material
and angular momentum and ensure an axisymmetric rotation pattern in the innermost 
disk area.

In the following, we denote the internal boundary by the term {\em sink}.
The sink geometry is a cylinder of unity radius $r_{\rm i} = 1.0$ and height $h_{\rm s}$.
Typically, $h_{\rm s} = 0.8$, and is resolved by 16 grid cells in height.
A sufficient grid resolution is required in order to resolve the cylinder by the Cartesian grid and 
to suppress effectively azimuthal asymmetries that could be induced by the rectangular grid cells.
We apply an equidistant resolution of
$(\Delta x=0.05, \Delta y =0.05, \Delta z=0.05)$ for the inner region of the grid, $-5.0 < (x,y,z) < 5.0$,
while for the rest of the domain a stretched grid is applied.
Thus, the circumference of the sink cylinder is resolved with about 125 grid elements.

One of the essential tasks for the model setup is to consistently prescribe a boundary condition for
the velocity components for the inner disk boundary.
Since we are using a Cartesian grid, both the accretion velocity and the rotational velocity are
interrelated with the $v_{\rm x}$ and $v_{\rm y}$ velocity components, and not easy to 
disentangle - adding numerical complexity when defining the boundary conditions.
We have therefore developed a set of boundary conditions that allow for an axisymmetric evolution 
of the inner region.

In Appendix \ref{sec:BC} we will discuss these boundary conditions in detail.

\subsection{Magnetic diffusivity}
Considering {\em resistive} MHD is essential for jet launching simulations.
Firstly, accretion of disk material across a large-scale magnetic field threading the disk plane 
perpendicular is only possible if that matter can diffuse across the field.
For a sufficiently long time evolution of the simulation, an equilibrium state will be
reached between inward advection of magnetic flux along the disk and outward diffusion
(see e.g. \citealt{2012ApJ...757...65S}).
Secondly, jet launching is a consequence from a re-distribution of matter across 
the magnetic field, and is therefore essentially influenced by magnetic diffusivity.

In our previous works, we have presented a detailed investigation about how the dynamics of the
accretion-ejection structure - such as the corresponding mass fluxes, jet rotation, or 
propagation speed - depends on the magnetic diffusivity profile and magnitude 
\citep{2012ApJ...757...65S, 2013ApJ...774...12F}.

In this paper we apply a magnetic diffusivity $\eta(r,z) \propto h_i(r,z)$ constant in 
time with thefollowing vertical profiles $h_i(r,z)$. 
Two different Gaussian profiles are applied,
\begin{eqnarray}
h_2(r,z) & = & \exp \left(-2\frac{z^2}{H^2}\right) \left(1 + \frac{0.1}{\exp(1-r)} \right) \nonumber\\
h_3(r,z) & = & \exp \left(-2\frac{z^2}{H^2}\right).
\end{eqnarray}
with the disk thermal scale height $H$.
However, we found that such diffusivity profiles may lead to instabilities in the 3D evolution 
of the system.

The most stable and smooth evolution of the accretion-ejection structure we observed when
applying a constant background diffusivity (as e.g. applied by \citealt{2003A&A...398..825V}).
Thus, for our reference run, a background value for the magnetic diffusivity was specified inside 
the disk and for the nearby disk corona,
\begin{eqnarray}
h_1(r,z) & = & \eta_0 \quad z<10,
\end{eqnarray}
while for the rest of the grid ideal MHD was assumed.

\section{A 3D gravitational potential}
\label{sec:3dGR}
In this section, we discuss the 3D non-axisymmetric gravitational potential that we apply in our simulations 
of jet launching on binary system.
For the purpose of this paper, we assume that both stars (respectively central objects) are sufficiently 
close, so that a 3D non-axisymmetric potential must be considered for the jet source. 
On the other hand, so far, we have neglected further details such as time evolution of the 3D geometry 
of the potential due to orbital motion or a mass exchange between the stars.

The effective gravitational potential for a binary system is given by the Roche potential,
\begin{eqnarray}
\Phi & = & -\frac{GM_{\rm p}}{|\vec r|}-\frac{G M_{\rm s}}{|\vec r-\vec r_{\rm s}|}-\frac{1}{2}\Omega^2 |\vec r-\vec r_{\rm CM}|^2, \\
\Omega & = & \sqrt{\frac{G({M_{\rm p} + M_{\rm s})}}{| D+ h|^3}}
    \nonumber
  \end{eqnarray}
where $r_{\rm p}$, $r_{\rm s}$ and $r_{\rm CM}$ 
denote the positions of the primary,
the secondary and the center of mass of the binary, respectively.
The stellar masses are denoted by $M_{\rm p}$ (primary) and $M_{\rm s}$ (secondary),
while $\Omega$ is the orbital angular velocity of the system.
The last term representing the centrifugal potential arises since
the reference frame of our simulations is not an inertial frame.

\begin{figure}
\centering
\includegraphics[width=1.\columnwidth]{\figurepath/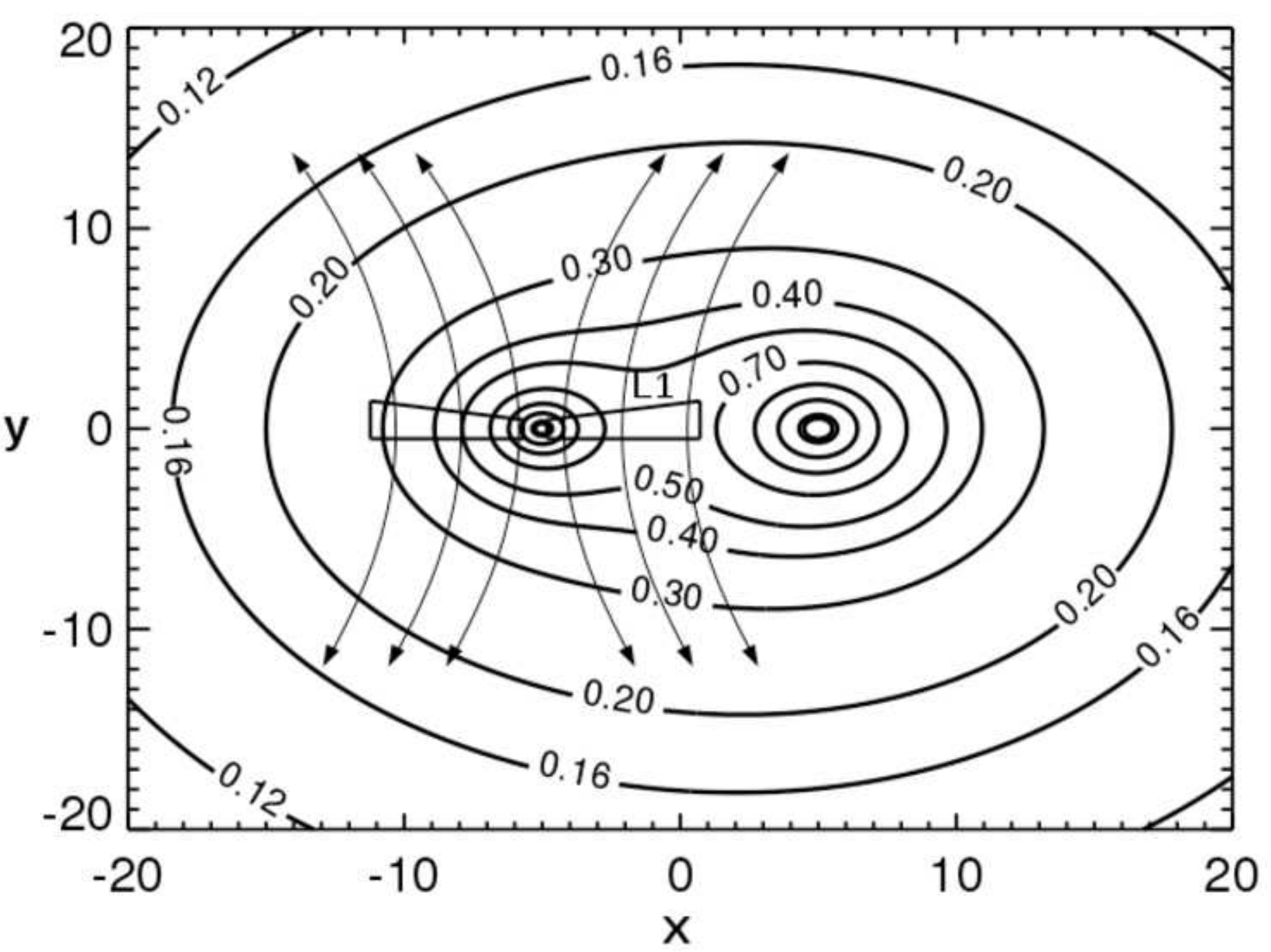}
\caption{Equipotential surfaces of the 3D Roche potential for the binary-disk-jet system.
Shown are the contours of the effective gravitational potential for a mass ratio of $q=M_{\rm s}/M_{\rm p}=2 $.
The binary separation is about 100 times the inner disk radius
(equivalent to 10\,AU for our scaling of protostellar jets).
The accretion disk located around the primary and the bipolar jets that are formed are indicated.}
\label{fig:binary_rochelobe}
\end{figure}

Figure \ref{fig:binary_rochelobe} indicates how the stars, the disk, and the jets are embedded in the 
Roche potential and the computational domain.
The figures shows true equipotential surfaces of the effective gravitational potential for a binary 
system with the mass ratio of $q=2$.
In particular, the inner Roche lobe is shown with the inner Lagrange point $L_1$ marked.
We have further indicated the flux surfaces of a collimating jet and counter jet launched from the
primary.
The jet material ejected first feels the gravity of the primary star only before it becomes influenced 
gravitationally by the secondary star along its path.
The jet superceeds the escape speed of the primary star shortly after launching and is
already super Alfv\'enic when leaving the Roche lobe of the primary star.
When it becomes affected by the secondary star, non-axisymmetric effects on the jet structure can
be expected.

Since our simulations are performed in code units, we are in principle able to scale our model 
results to different sources that may launch jets, applying proper length scales or magnetic field 
strengths.
We show typical parameters of binary sources in a brief observational summary in 
Table \ref{tbl:binary star-observation}.
In order to estimate the influence by a companion for the accretion-ejection evolution, 
it is useful to compare the various physical time scales in the binary star-disk-jet system, 
such as the orbital period of the system, the time scales for precession of warping instabilities,
and the dynamical time scales of the outflow.

The orbital period of the stars is
 \begin{eqnarray}
  T_{\rm orbit} & = & 2\pi\sqrt{\frac{(| D+ h|)^3}{G (M_{\rm p} + M_{\rm s})}}\\
 & =  & 60
 \left(1+k \right)^{-1/2}
  \left(\frac{(| D+ h|)}{15 \rm AU}\right)^{3/2}
  \left(\frac{M_{\rm p}}{M_\odot}\right)^{-1/2} {\rm yrs}, \nonumber
 \end{eqnarray}
where $q = M_{\rm s}/M_{\rm p}$ denotes the secondary-to-primary mass ratio
and $| D + h|$ is the separation between the two stars.

The dynamical time step in the simulations is given by the length unit (inner disk radius $r_{\rm i}$) and the
orbital velocity at the inner disk radius ${v_{\rm Kep}}$,
\begin{eqnarray}
T_{\rm in} & = & \frac{r_{\rm i}}{v_{\rm Kep}}
            = r_{\rm i}\left(\frac{GM_{\rm p}}{r_{\rm i}}\right)^{-1/2} \\
            & = & 1.8 \left(\frac{r_{\rm i}}{0.1\rm AU}\right)^{3/2}
             \left(\frac{M_{\rm p}}{M_\odot}\right)^{-1/2} {\rm days }\,\, {\rm (YSO)} \nonumber \\
                 & = & 0.5 \left(\frac{r_{\rm i}}{10R_{\rm S}}\right)^{3/2}
             \left(\frac{M_{\rm p}}{10^8 M_\odot}\right) {\rm days } \,\, {\rm (AGN)}. \nonumber
\end{eqnarray}
Here, we assume typical parameters for young stars (YSO) and active galactic nuclei (AGN).
For YSO the inner disk radius is $r_{\rm i} \simeq {0.1\rm AU}$ while for AGN $r_{\rm i} \simeq {10R_{\rm S}}$ 
(where $R_{\rm S}$ is the black hole Schwarzschild radius).

The typical running time of our simulations is 3000-5500 dynamical times, corresponding to 15-30 years in 
case of YSOs and  5-10 years in case of AGN.

Other essential time scales are those for disk wobbling and disk precession in a binary system for which 
disk is misaligned with the orbital plane of the binary. 
It has been argued that the period of outer disk plane to wobble is half of the orbital period of the binary, 
$T_{\rm orbit}/2$ \citep{1982ApJ...260..780K, 2000MNRAS.317..773B}.
Furthermore, \citet{2000MNRAS.317..773B} have shown that tidal forces will cause a precession of the disk axis
with a period of the order $T_{\rm P} \simeq 20 T_{\rm orbit}$.
In addition, these authors estimated that the characteristic time-scale disk realignment along the orbital 
plane owing to dissipation is of the order of the viscous evolution time-scale, i.e. the order of 100 
precession periods.

For the purpose of this paper we do not consider the motion of the binary along its orbit for the 
effective gravitational potential, and assume it as constant in time.
Depending on the simulation time scales, this assumption is reasonable, as long as the binary separation is 
sufficiently large.
The typical separation in our simulations is chosen as to $300\,R_{\rm i}$, corresponding to about $ 30\,{\rm AU}$.
On the other hand, for large values of binary separation the corresponding time scale for precession would 
be so long that signatures of disk precession would not be expected during typical simulation time scales.

The distance of $L_1$ from the less massive star\footnote{Note that we have defined the primary as the star hosting
the jet launching disk, and not as that star with the higher mass}, $L_1$, 
are approximately given by the fitting formula by Plavec \& Kratochvil
\begin{equation}
   \label{Roche}
  L_1 = D (0.5+0.227 \log q)
    \nonumber
\end{equation}
while towards the more massive star the distance of $L_1$ is $D (0.5-0.227 \log q)$
(see e.g. \citealt{1992apa..book.....F, 1997ASSL..216.....C}).
If $L_1$ is located close to the accretion disk that we initially prescribe, or even inside the disk,
some of disk material will be transferred outwards (from the computational domain) towards the secondary.
The size of the accretion disk will then be determined by the size of the Roche lobe.
Also, the stability of the initial Keplerian disk will be affected by the Roche potential.

For the purpose of this paper, we have chosen different stellar mass ratios and binary separations.
For a wide binary separation, the $L_1$ is located outside the computational box and allows for a long-term
evolution simulation, just because the disk mass can be maintained for longer time.
However, in order to be able to trace the tidal effects that would otherwise happen only on a much larger 
time scales, we will also present one simulation with a small separation between the stars (see section \ref{extreme case}).
 
In the following we will present our simulation results by {\em snapshots} observed in the {\em reference 
frame of the primary.}
In order to visualize our results as sky maps - as they would be seen by a terrestrial observer - one would 
need to project simulation the data onto the plane of the sky, 
considering a time dependent coordinate transformation into the center of mass coordinate system, and thereby 
considering the orbital motion of the jet launching primary.
This is beyond the scope of the present paper that is devoted purely to the 3D dynamics of jet launching.

Here we summarize the essentials of our the model setup.
\begin{itemize}
\item[(1)] The origin of the coordinate system is centered at the primary star that is surrounded by an 
  accretion disk forming bipolar jets. We test our model setup with 3D simulations of jet launching from 
  a single star accretion disk.
\item[(2)] Our first focus is on studying the jet launching in a binary system with a separation sufficiently
  wide so that no mass transfer happens. 
  In this case the orbital time scale is larger than the time scale of our simulation,
  and the center of mass is located outside the simulation box in most cases.
\item[(3)] The jet launching area is located inside the inner Roche lobe of the primary. 
  Once the outflow is formed, it propagates beyond the Roche lobe. 
  The outflow is then influenced by the 3D gravitational potential, and the outflow propagation 
  will deviate from a straight, axial motion.
\item[(4)] In order to investigate the onset of disk precession and subsequent jet precession, we also
  run a simulation with a small binary separation, such that the precession time scale is comparable
   to the simulation time scale.
\end{itemize}
Applying this model setup, we will present the first ever results of 3D MHD simulations of bipolar jet 
launching from magnetized accretion disk.

\section{Axisymmetric jet launching in 3D}
We first discuss our reference run {\em scase2} that allows us to test our 3D model approach and
in particular to examine the symmetry and the stability of the setup.

The setup of the reference run considers the gravitational potential of a single star, the disk surrounding 
that star, and an initial coronal structure extending from the disk surface into both hemispheres.
We have performed a series of parameter runs (see Table \ref{tbl:3D runs}) in particular applying
different magnetic diffusivity models.

\subsection{Magnetic diffusivity and hemispheric symmetry}
By investigating different prescriptions for the magnetic diffusivity profiles, we recognized that some 
of them may directly affect the symmetry of the bipolar jet-disk structure - in spite of the symmetric 
and well-tested inner boundary condition.
This is in contradiction to our recent axisymmetric simulations
\citep{2012ApJ...757...65S, 2013ApJ...774...12F} for which the bipolar
symmetry was well kept for several 1000 rotations for various model setups. 
We have checked this carefully, without, however, coming to a finite conclusion.
We find that by increasing the magnetic flux the asymmetric evolution begins earlier in time and also closer 
to the internal boundary.
We conclude that the magnetic field may play a significant role in amplifying an asymmetric perturbation.
The magnetic diffusivity directly influences the induction of toroidal electric currents and the re-distribution 
of the magnetic field by Ampere's law.
Tiny numerical differences caused by the rather low resolution of the exponential profile of diffusivity may 
therefore be responsible for introducing a slight offset in the magnetic field structure in both hemispheres.
The grid resolutions in 3D is 20 grid cells per length unit, or 2 grid cells per disk initial thermal scale 
height $H$.

Further disturbance of symmetry may arise from reconnection events that may introduce a kind of stochasticity 
(as reconnection cannot be resolved on our grid).
Reconnection in the diffusive disk may locally alter the electric current distribution, and, as a 
consequence, also the Lorentz force that is involved in launching the outflows.

In order to suppress any artificial jet asymmetry that may be triggered by artifacts induced by the
magnetic diffusivity profile, we decided to apply the background diffusivity approach $h_1$.
These runs apply a constant diffusivity distribution across the disk - for {\em scase1} we apply a 
constant background diffusivity over the whole domain, while for {\em scase2} a constant diffusivity 
was defined for the region $|z| \le 10$.
As a result, for the two runs {\em scase1} and {\em scase2} the bipolar symmetry for jet and the counter jet 
was very well kept for 600 rotations (4000 dynamical time steps). 

We choose simulation {\em scase2} as reference for our 3D simulations, since a magnetic diffusivity that is
confined to the the disk / jet launching area seems to be closer to reality (of a stratified disk) and is
better comparable to literature papers that typically assume an exponential diffusivity profile vertically.

\subsection{General evolution of accretion-ejection in 3D}
In the following we will discuss the evolution of our reference run {\em scase2}.
Before going into details, it is interesting to see a fully 3D presentation of our simulation result.

\begin{figure}
\centering
\includegraphics[width=1.\columnwidth]{\figurepath/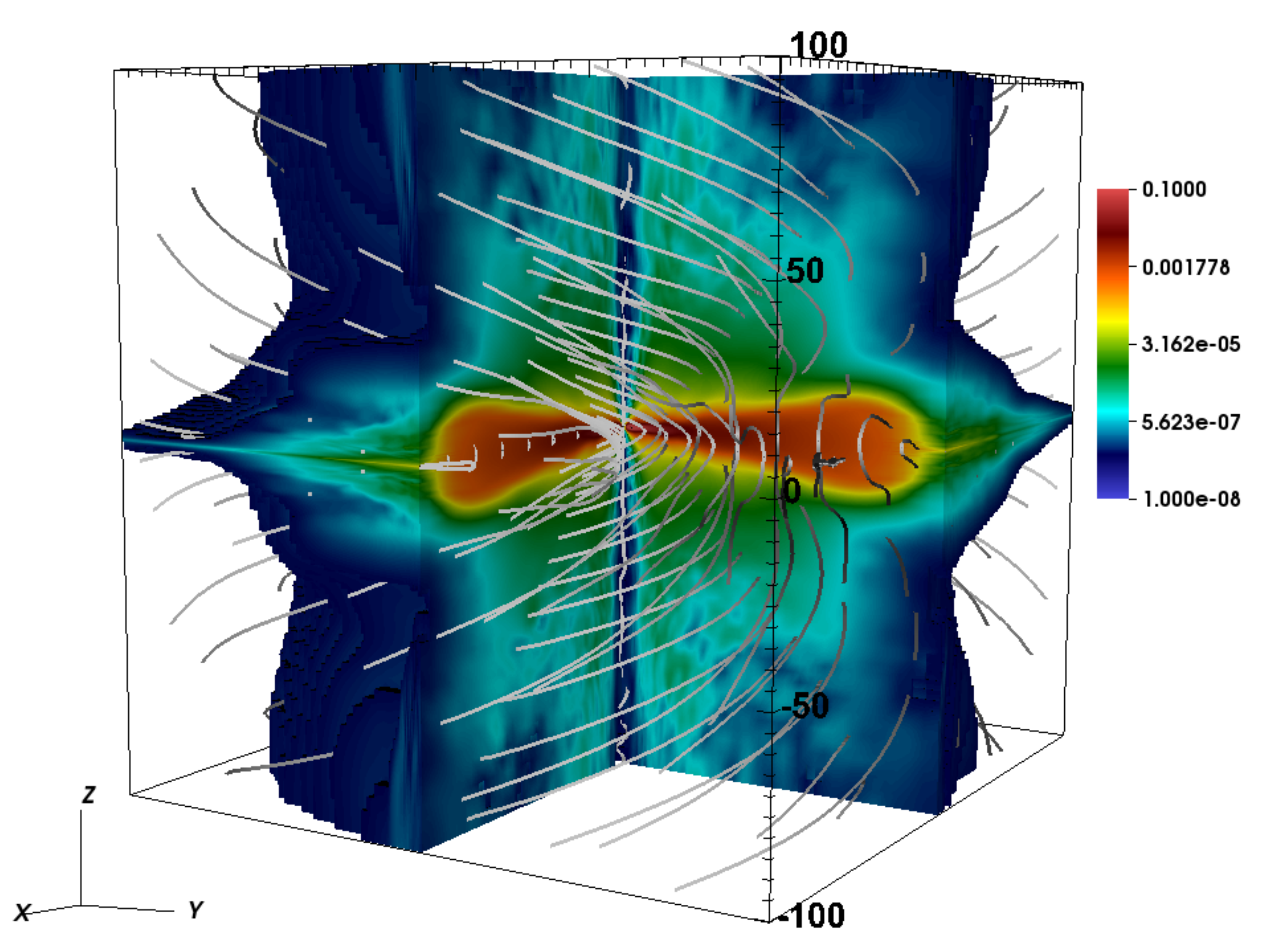}
\caption{ Mass density in 3D. Shown is the cut of
the mass density distribution in three dimensions for the reference run ``scase2'', at time 500.
The stream lines of the magnetic field are displayed in gray.}
\label{fig:3D_testrun}
\end{figure}

Figure \ref{fig:3D_testrun} shows a rendering of the mass density distribution in three 
dimensions for the reference run {\em scase2} at time $t=500$.
To allow for a proper visualization of the internal structure of the disk-jet, we have 
applied a threshold density of $10^{-7}$, i.e. the surrounding corona is invisible.
The gray lines follow the magnetic field lines.
The 3D visualization in Fig.~\ref{fig:3D_testrun} displays the evolved disk/jet system simultaneously. 
It shows that bipolar jets that are symmetrically formed from the magnetized disk.
The disk (orange colors) is dense, the jets are dilute (green-blue colors)
and have formed at this stage ($t=500$) from about half of the disk surface.

\begin{figure*}
\centering
\includegraphics[width=13.5cm]{\figurepath/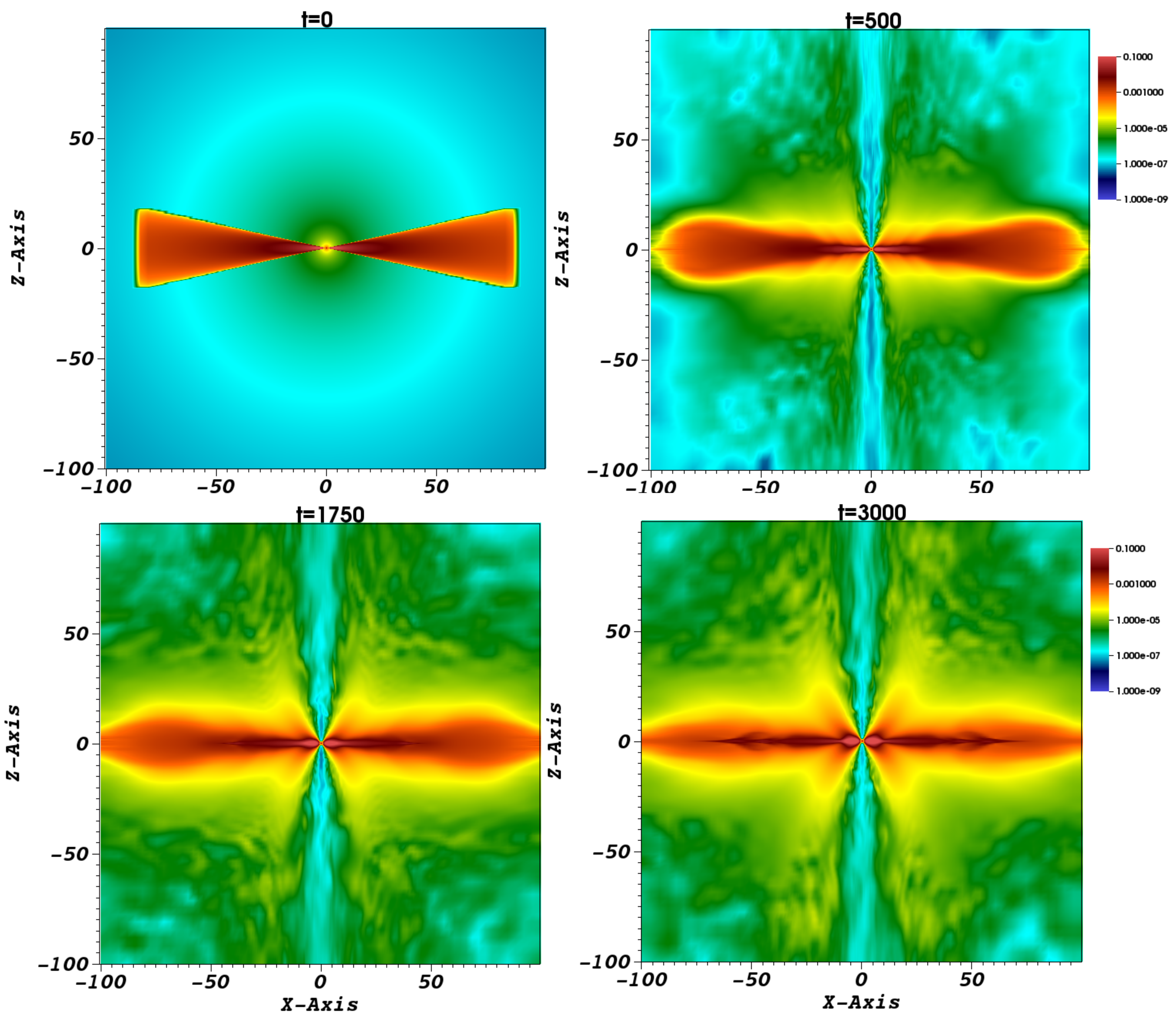}
\caption{Time evolution of reference run {\em scase2} following an axisymmetric setup in 3D.
Two dimensional slices of the mass density in the $x$-$z$ plane, at times $t=0, 500,1750,3000$.}
\label{fig:fct9_rho_xz}
\includegraphics[width=13.5cm]{\figurepath/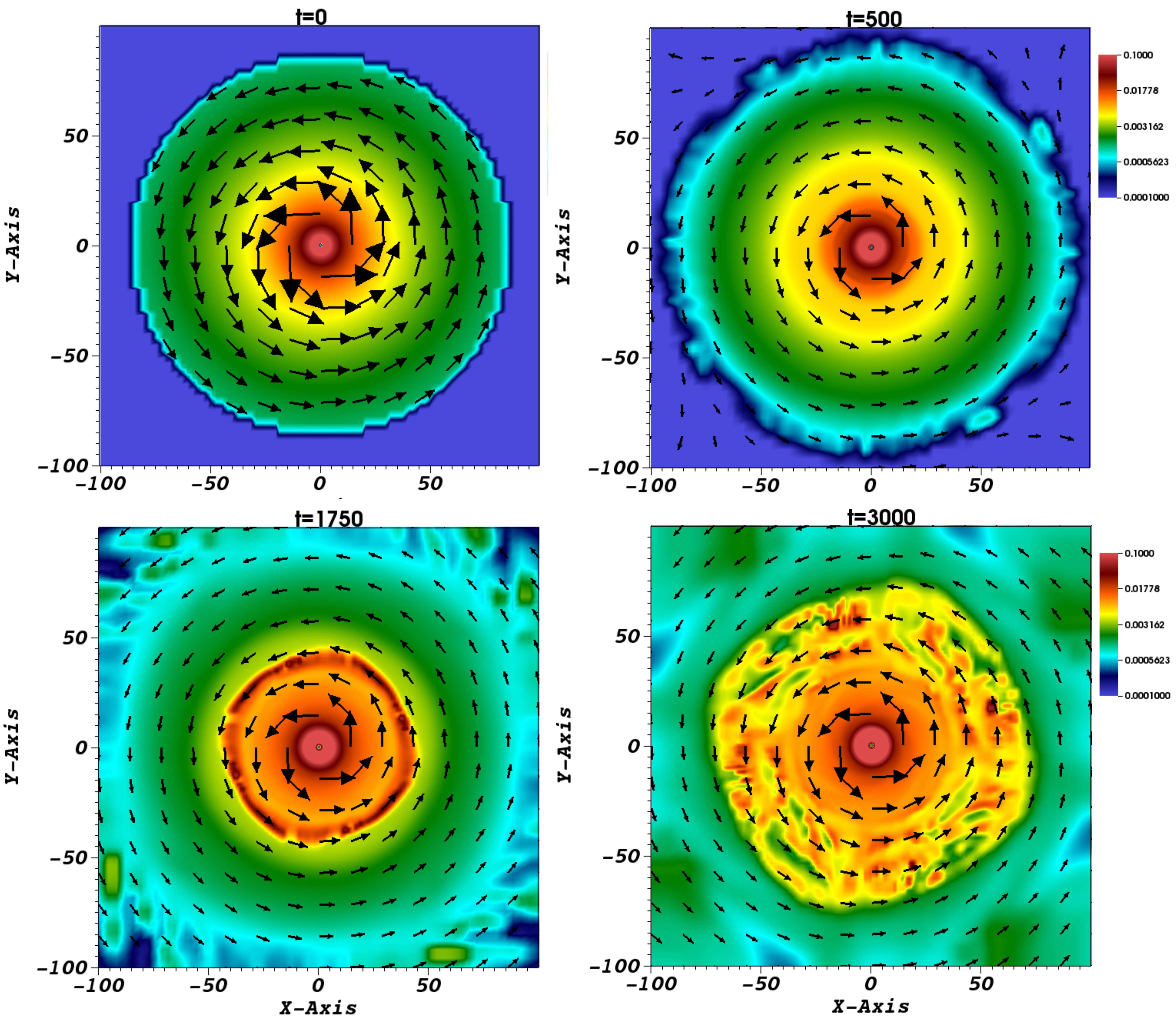}
\caption{Time evolution of the disk in reference run {\em scase2}
following an axisymmetric setup in 3D.
Two dimensional slices of the mass density in the $x$-$y$ plane (mid-plane), at times $t=0, 500,1750,3000$.
The arrows indicate to the velocity vector of the disk.}
\label{fig:fct9_rho_xy}
\end{figure*}

We further demonstrate the quality of our the model setup, in particular the symmetry of the launching
process, by showing the time evolution of the mass density in Figure \ref{fig:fct9_rho_xz} 
(from aside) and Figure \ref{fig:fct9_rho_xy} (from top).
The different panels in Figure \ref{fig:fct9_rho_xz} show the slices of the 3D density distribution,
typically in the $y=0$ plane containing the rotation axis (in z-direction).
The initial time step with the hydrostatic corona is not shown.
We run this simulation for 4000 dynamical time steps until the disk mass loss due to accretion and ejection 
has become substantial.

Naturally, we expect the small-scale structure to be different from axisymmetric simulations
\citep{2007A&A...469..811Z, 2009MNRAS.400..820T, 2012ApJ...757...65S},
given the somewhat lower numerical resolution and the treatment of a third dimension.
For example, we observe that a layer of perturbed material appears at $t\simeq 500$ along the inner part of the 
outflow.
However, the overall large-scale disk-outflow structure shows a well-kept right-left (rotational) symmetry 
and also a good bipolar (hemispheric) symmetry of the outflow for about 600 inner disk rotations,
in general approving the quality of our model setup.

Since we treat a rotating system in Cartesian coordinates, it is interesting to check in more detail the
rotational symmetry of the accretion-ejection structure.
In Figure \ref{fig:fct9_rho_xy}, we show the evolution of the mass density in the  equatorial plane 
for the dynamical time steps $t= 500, 1500, 2000, 3000$.
This slice traces the structure of the accretion disk.
We see that the mass density indeed maintains an axisymmetric distribution until late evolutionary stages.

However, small-scale density fluctuations appear at intermediate radii at around $t\simeq 1500$.
They arise at a radius $r=40$ and then extend to larger radii for later times.
At $t=3000$ a ring of density fluctuations exist extending from $r=30\,R_{\rm i}$ to $r=60\,R_{\rm i}$.
This ring is not fully concentric anymore, but has adopted a slightly rectangular shape, as the outer layers 
of the disk are affected by the shape of the computational box.
The amplitude of the fluctuations grows in time and may finally reach values of up to 50\%.
These fluctuations are visible in different physical variables, such as mass density, gas pressure, velocity.
The nature of this feature is not yet clear to us and definitely deserves a detailed investigation.
However, since it is not closely connected to the launching of the inner jet, we refer such a study to a 
future paper.
So far, we suggest that these fluctuations may be caused by a magneto-rotational instability (MRI) working 
in these outer parts of the accretion disk.
Here, the magnetic field is rather weak and the grid resolution per disk height is sufficiently high in 
order to resolve the Alfv\'en wavelength, and thus the MRI. 
Furthermore, the magnetic diffusivity is lower compared to the inner disk radii.

\subsection{Mass flux evolution of the reference run}
In order to further test the symmetry and the compatibility of our 3D model setup,
it is useful to follow the evolution of the mass flux distribution away from the disk.
To measure the mass flux, we adopt a rectangular box of size defined by $x,y=10$ and $z=3$ 
around the origin and integrate the mass fluxes $\rho v_{\rm z}$ and $\rho v_{\rm r}$
across the corresponding surfaces.

\begin{figure}
\centering
\includegraphics[width=1\columnwidth]{\figurepath/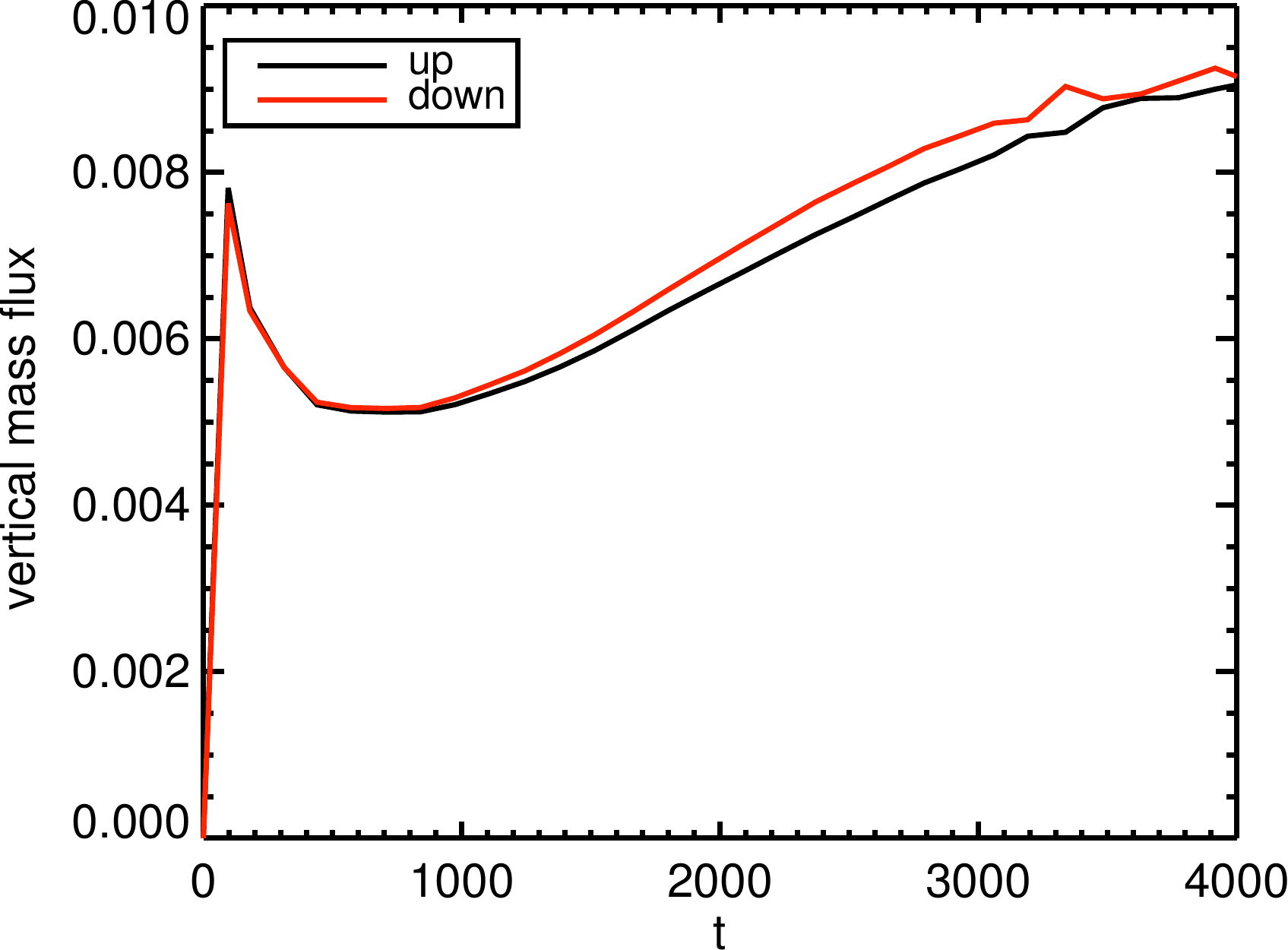}\\
\includegraphics[width=1\columnwidth]{\figurepath/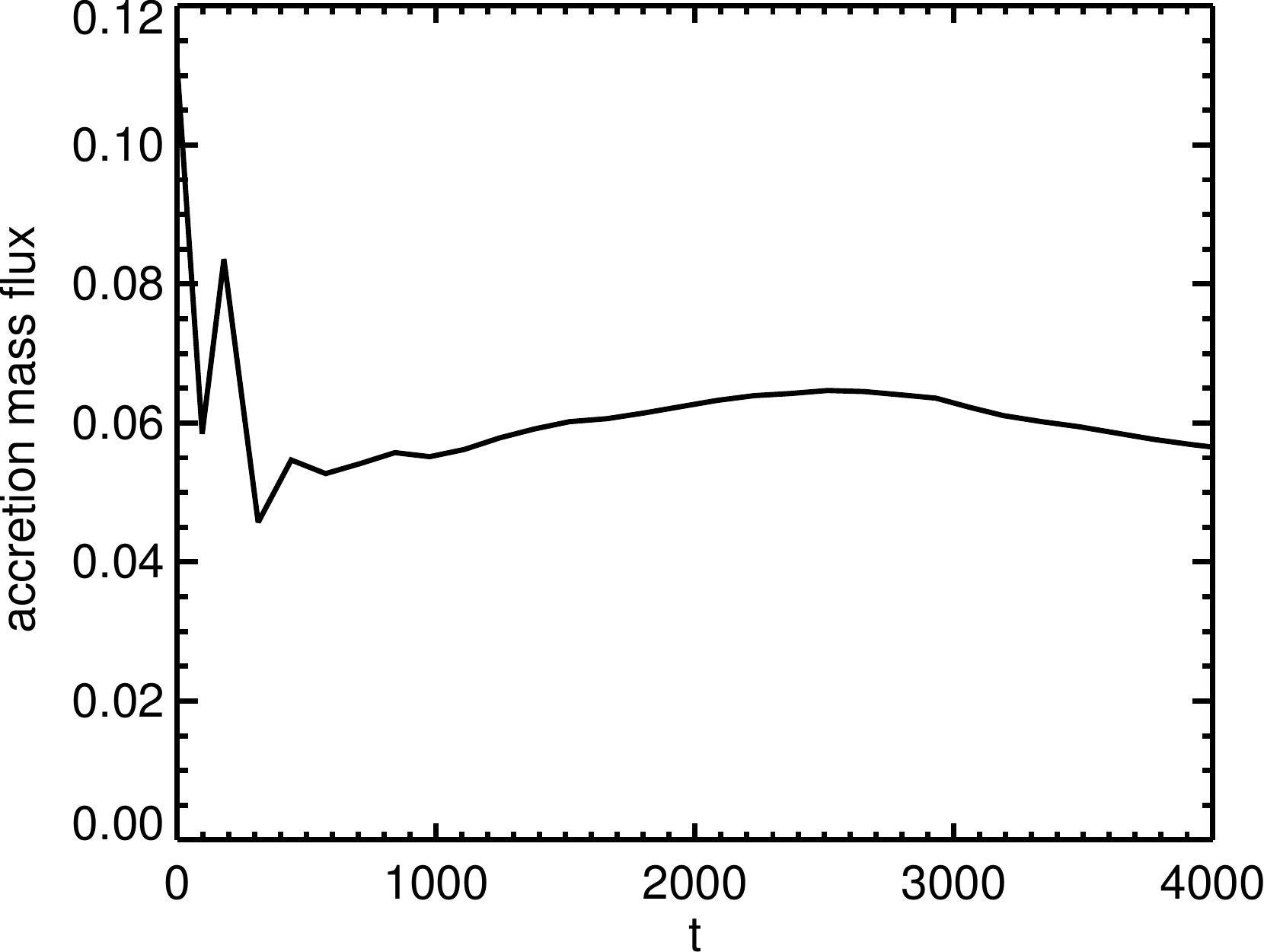}
\caption{Evolution of the mass fluxes in the accretion-ejection system for the reference run {\em scase2}.
Shown are the mass flux in the outflows (left) into both hemispheres and the accretion rate (right).
The mass fluxes are integrated over a box of size $x,y=10$ and $z=3$.}
\label{fig:scase2_ejection}
\end{figure}

Figure \ref{fig:scase2_ejection} shows the time evolution of the mass fluxes measured for the reference 
run {\em scase2}.
On the top, we plot the evolution of the ejection rate into both hemispheres, while on the bottom
the evolution of the accretion rate is shown. 
Comparing the ejection rates into both hemispheres (Figure \ref{fig:scase2_ejection}, top), 
we see that there are basically the same (the difference is 5\%).

The ejection rate is increasing by about $30\%$  during the evolution 
while the accretion rate saturates after 500 dynamical time steps.
In code units, we find an accretion rate of $0.06$ and an ejection rate between 0.006 - 0.008.

This gives an ejection-accretion ration of about $0.1$, that is definitely lower than for the
axisymmetric simulations \citep{2012ApJ...757...65S}.

It seems that the evolution of the vertical mass flux saturates to a constant value at late stages of about 0.009.
We think that the most important reason why saturation is not reached (will never be reached), 
is because the underlying launching conditions slowly change in time. 
This is a natural outcome
of the small disk size applied in our simulations, that limits the mass reservoir for
accretion and ejection.
This is a natural outcome of the small disk size applied in our simulations, that limits the mass reservoir for
accretion and ejection.
Thus, in this sense, reference run does not reach a steady state, but a {"}quasi steady state{"}.

The rough numbers for the mass fluxes are, however, comparable with those obtained in the axisymmetric  
setup \citep{2012ApJ...757...65S}, where we found accretion rates of 0.015 and ejection rates of
0.008  (in code units) for the same control volume.
Clearly, the exact numbers depend on the further parameter choice such as magnetic field strength
or magnetic diffusivity.
In comparison, the outflow mass flux in 3D is of the same order as for the axisymmetric ejection
while for the accretion rate we obtain a higher value in 3D.
This is due to the fact that a different magnetic diffusivity profile is used for the 2D
and the 3D simulations, and also the grid resolution is different.
In particular the lower resolution - implying a somewhat higher numerical diffusivity - will both increase 
the accretion rate, but also increase the jet mass loading.
In addition, for the 3D setup we have applied a the lower magnetic field strength.
As a consequence, the mass flux ejected into the outflow is lower.

Overall, we find from the long term evolution of the jets launched in our 3D setup that both disks and
jets evolve into a stable and symmetric structure, confirming the quality of our 3D model setup.
Having approved our model setup, allows us to continue and further investigate non-axisymmetric effects
resulting from different physical situations.
In the next section, we perturb the symmetry of the initial disk-jet structure by a companion star in 
a binary system.

\section{A binary system - jet inclination and disk precession}
In this section, we present results of simulations considering a 3D effective gravitational
potential of a binary system.

In this setup, the vertical separation of the secondary from the initial disk mid-plane, parametrized 
by $h$, implies that the accretion disk is misaligned with respect to the orbital plane
(see Figure \ref{fig:binary_structure}).

The mass ratio and the binary separation are the most significant parameters that determine the 
characteristics of the binary system, such as the position of the Lagrange points or the kinematic 
time scales of the system - the larger the separation, the larger time scales are.

On the other hand, the numerical simulation is substantially constrained by (disk) mass reservoir 
available for accretion and ejection.
Since the disk continuously loses its mass via the internal boundary and the outflow, we can not run a 
simulation too long.
This holds in particular for close binary systems since the disk size is then limited by the Roche lobe.
On the other hand, some 3D tidal effects of the binary star-disk-jet evolution will be visible only on 
comparatively long time scales.

In the following we first discuss simulations applying a rather wide binary separation {\em bcase1}, 
before we present results of an extreme parameter set {\em bcase2} that clearly exhibits tidal effects
from the binary system.

\subsection{Global outflow asymmetry beyond the Roche lobe}
\label{binary case}
In this section, we discuss simulations based on a parameter choice that does not allow to observe 
tidal effects on the dynamical evolution of the disk-jets system - just because the time scales of 
those are much longer than our setup allows.

Nevertheless, even for a binary system with a rather wide binary separation, we may study 3D 
effects of jet formation on various scales.
The jet launching area is located well within the inner Roche lobe, and the
jet is formed in axisymmetry.
The situation changes when the jet leaves the Roche lobe, since the jet propagation is then
affected by the gravity of the secondary.
As a result, the jet motion may deviate from the original direction of propagation along the 
rotational axis of the primary and the accretion disk.

\begin{figure*}
\centering
\includegraphics[width=18cm]{\figurepath/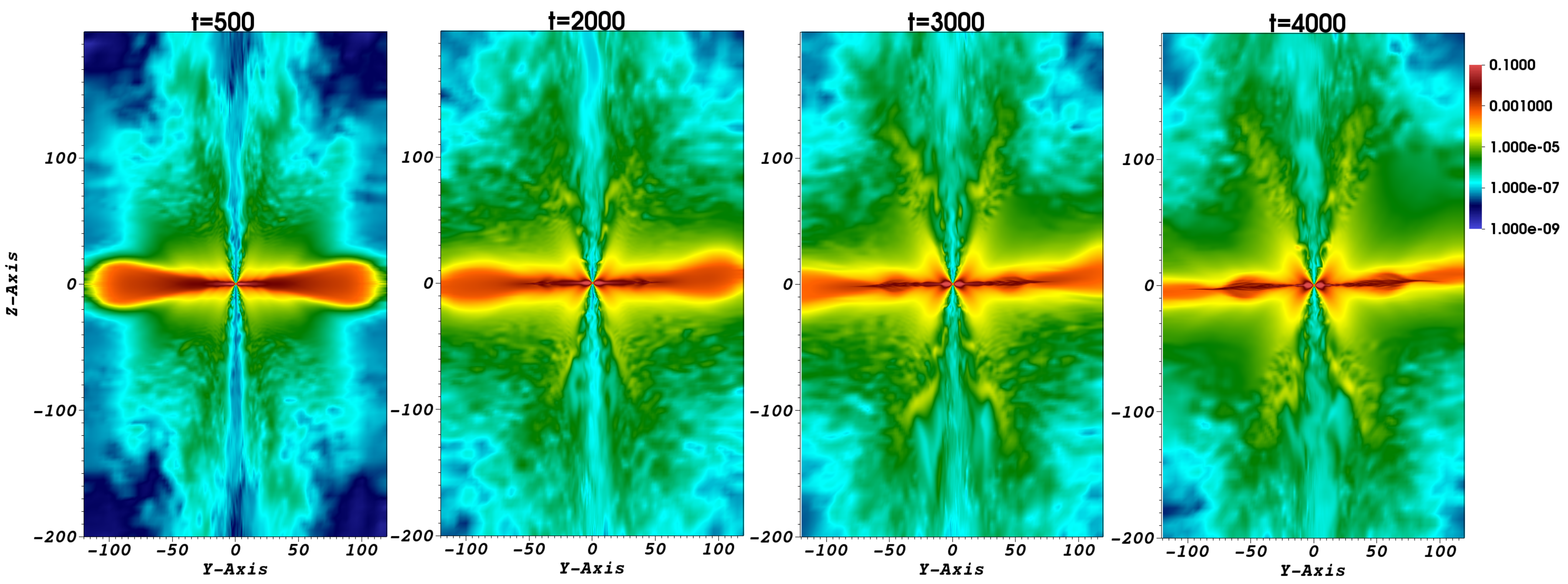}
\caption{Binary star-disk evolution. Shown are the two dimensional slices of the mass density
done for the binary star-disk-jet run, {\em bcase1}.
The slices show the mass density in $y-z$ plane, for times 500, 2000, 3000, 4000.}
\label{fig:bcase1_rho_xzyz}
\end{figure*}

This evolution can be seen in Figure \ref{fig:bcase1_rho_xzyz} where we show two-dimensional slices
in the $y$-$z$ plane
of the mass density for simulation run {\em bcase1}, for times $t= 500, 2000, 3000, 4000$. 
We observe a deviation from the straight propagation along the initial rotational axis, in particular
after $t=4000$.

In addition, we see that the accretion disk is not anymore aligned with 
the initial disk mid-plane (the equatorial plane).
Instead, it appears that the disk tends align with the orbital plane.
However, as mentioned already, the characteristic time-scale for an alignment of the disk
with the orbital plane is of the order of 100 precession periods \citep{2000MNRAS.317..773B}. 
Therefore, the change in the disk alignment may suggests that we observe the very initial stages
of disk precession (see below).

We further observe that the disk expands beyond the initial outer radius of the disk.
Concerning the global disk structure, two further effects can be seen fluctuations and a bump in the 
overall disk.
The fluctuations are observed as deviations from a smooth disk structure and are
seen in various disk variables such as mass density, pressure, magnetic field strength and velocity.
They form outside $r = 30$ and extend to larger radii.
As discussed above, we think that these fluctuations are signatures of the magneto-rotational 
instability in the disk
\citep{2007A&A...476.1123F, 2010A&A...516A..26F, 2013A&A...550A..61L, 2013ApJ...775..103U}.
We will follow-up this idea in a future work, however, it is interesting to note
that we see a disk wind driven also from these perturbed disk areas.

\begin{figure}
\centering
\includegraphics[width=\columnwidth]{\figurepath/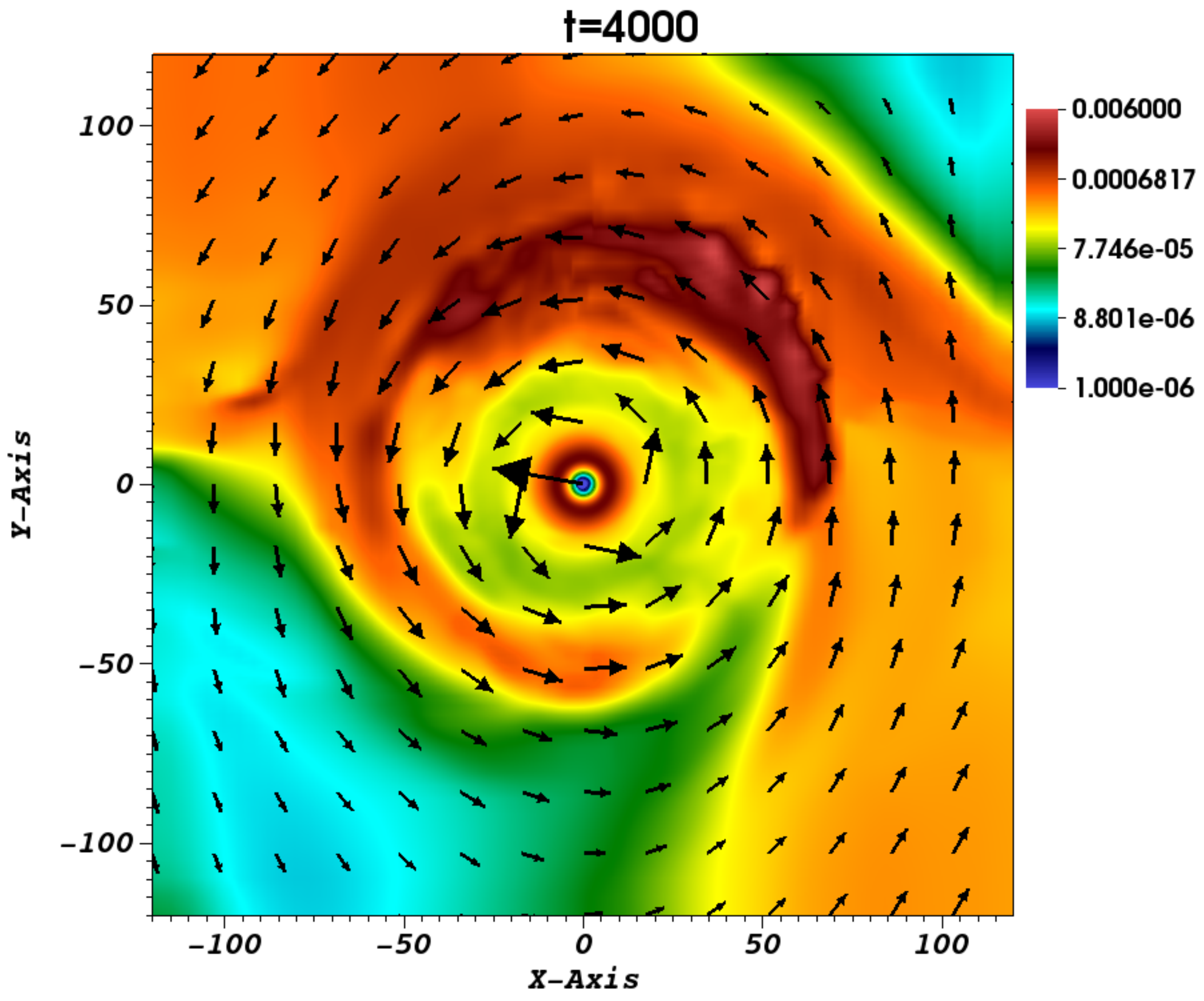}
\caption{Non-axisymmetric evolution of the launching area.  
Shown is a slice of the mass density for simulation bcase1 considering a binary system
in a plane parallel to the $x$-$y$ plane, at height $z=5$ and for time 4000.
The arrows indicate the velocity vector (also measured at height $z=5$).}
\label{fig:bcase1_hump}
\end{figure}

The bump is a large scale 3d asymmetry in the disk structure and is predominantly seen in the mass 
density and the mass flux profiles  ($(\rho v)$).
Figure \ref{fig:bcase1_hump} shows a slice of the mass density parallel to the $x$-$y$ plane at $z=5$ 
for simulation {\em bcase1} considering a binary system.
In this slice the bump is distinguishable as a high density area (red color).

Other disk variables, such as pressure, velocity, or magnetic field do not exhibit this feature.
The bump is built up along the direction towards the companion star, but then continues to build-up a
high density ring structure as the material is orbiting around the primary.

We interpret the formation of the bump as a first signature of disk warping.

\subsection{Precession of jet nozzle and jet?}
\label{extreme case}
Above we have mentioned the limitations of our simulations by the available mass reservoir for disk 
accretion and jet ejection, and the time scales of the time evolution of the binary system. 
Here, we discuss simulation {\em bcase2} that nicely demonstrates the 3D effects of jet launching
that may be caused by tidal effects of a companion star.

In order to be able to observe 3D tidal effects in our simulations, we have applied an extreme parameter 
setup, essentially governed by a small binary separation.
Here, the separation between two stars is only $\sqrt{200^2+60^2}\,r_{\rm i}=209\,r_{\rm i}$ 
(or 21\,AU for protostallar scaling) and the mass ratio is unity, $q=1$.
Therefore, possible tidal effects such as disk warping or jet precession are expected to happen 
much faster in this system.
Note that due to the smaller separation, the inner Lagrange point $L_1$ is now located {\em inside}
the simulation box and also inside the initial accretion disk, namely at $(x, y, z)=(100,0,30)$.

\begin{figure*}
\centering
\includegraphics[width=18cm]{\figurepath/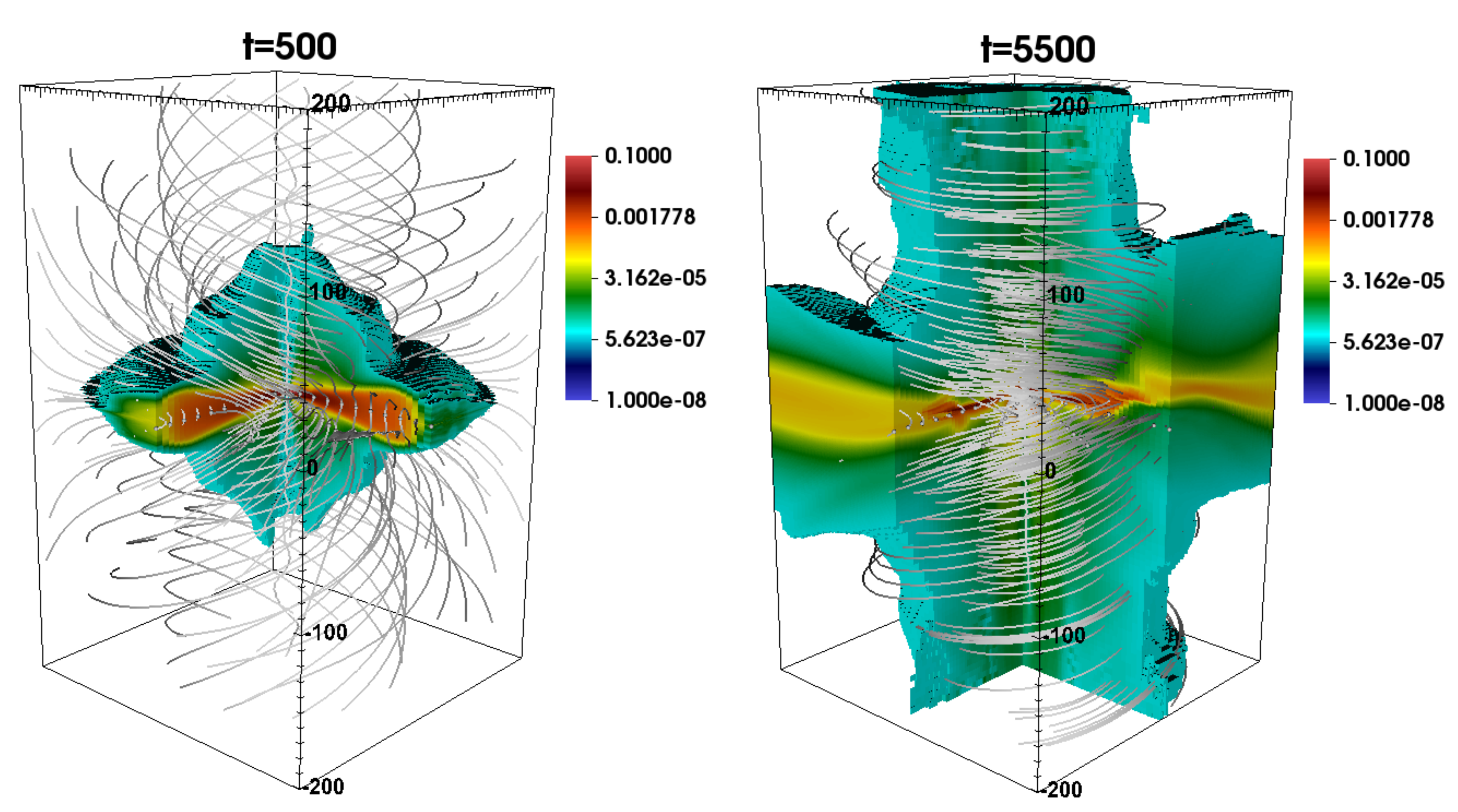}
\caption{3D evolution of disk-jet structure. 
Shown is the distribution of the mass density (color coding) and magnetic field lines (grey lines) 
for simulation run {\em bcase2} assuming a binary system with rather close separation.
A density threshold  about $10^{-6}$ is applied in order to visualize the inner disk-jet structure.
The  slices of the 3D distribution are shown for the dynamical time steps $t= 500, 5500$.}
\label{bcase2_warping}
\end{figure*}

We will now discuss run {\em bcase2} in detail and compare it with our other simulation runs considering 
binary systems (see Table \ref{tbl:3D runs}).
Figure \ref{bcase2_warping} shows the time evolution of the mass density in a 3D rendering.
In Figure \ref{bcase2_warping} a density threshold of $10^{-6}$ is applied for the rendering
in order to show the disk-jet evolution inside the surrounding corona.

We observe that, similar to the simulation discussed before, in {\em bcase1} a kind of inflation or flaring
of the accretion disk takes place beyond the $L_1$ point that is initiated at time scales $t> 1000$.
This inflation is directed towards the secondary and is seen in particular in the upper hemisphere
(that is closer to the companion star).
The Roche lobe overflow is seen in the radial velocity profile of the disk with positive radial velocities 
close to the inner Lagrange point $L_1$ for simulation {\em bcase2}. 
We also observe that the initial accretion disk dissolves beyond L1 and no disks exist beyond 
the Roche lobe for late evolutionary times.

Furthermore, we observe a similar fluctuation pattern as in simulation {\em bcase1} and also a bump
in disk mass indicating localized accumulation of mass in the outer disk.

The signature of the jet inclination is more distinct in this case.
This may be expected as the binary effects are larger now due to the smaller binary separation.
The bipolar jets launched from the disk first follow a direction along the $z$-axis before
they deviate from the initial propagation direction.

The structure and the alignment of the accretion disk changes.  
We recognize that the disk becomes more and more misaligned of with respect to the initial mid-plane.
We believe that this effect indicates the onset of {\em disk precession}, although we are not able
to observe a full precession cycle during the run time of our simulations.
As a consequence, also the jets become launched in a different direction.
This effect is in particular observable in the upper hemisphere that is closer to secondary and in 
which the material is stronger affected by the corresponding forces of the binary system.
Apart from the intrinsic change of the jet launching direction, also jet inclination happens - predominantly 
above $z=50$, just outside the Roche lobe of the primary in case {\em bcase2}.

\begin{figure}
\centering
\includegraphics[width=1.\columnwidth]{\figurepath/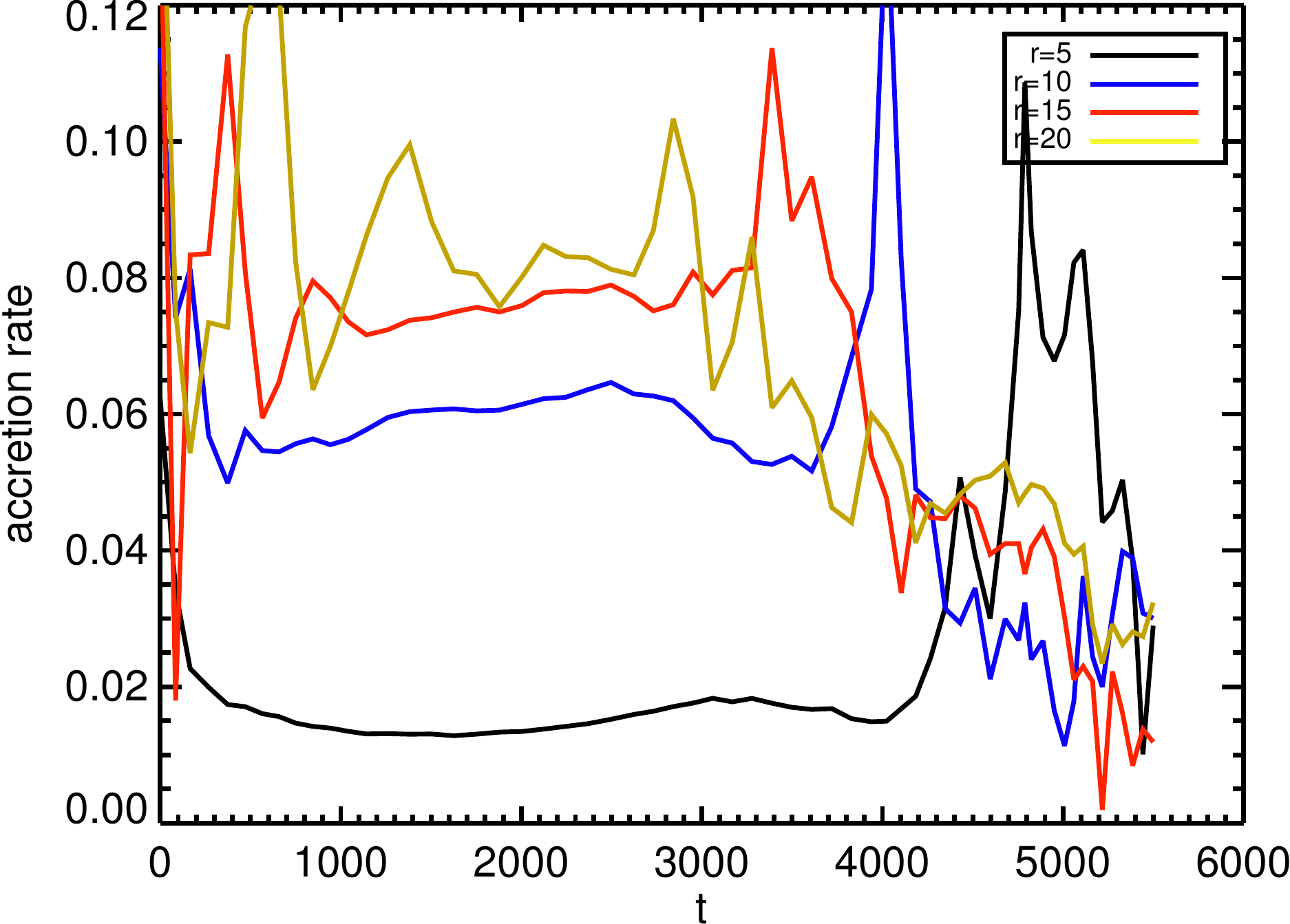}
\caption{Accretion mass flux evolution for simulation run {\em bcase2}. 
Shown is the evolution of the accretion rate measured at radii $x,y = 5, 10, 15, 20$. 
and height $z = 3H$ where $H=H(x,y=10), ..., H(x,y=20)$ is the thermal scale height of the disk,
at this radius.
}
\label{bcase2_accretion}
\end{figure}

The formation of a warped disk structure subsequently affects the launching process.
Therefore, we expect to see the footprints of the disk warps in the mass flux evolution, as well.
In order to confirm this effect, we measure the accretion rate at 4 different radii of the disk, 
$x,y = 5, 10, 15, 20$ (see Figure \ref{bcase2_accretion}).
We measure the accretion rate at each radius by integrating a rectangular box with height of three thermal
scale heights $z=3H$ calculated at the outer box radius with $H = H(x,y=5), ..., H(x,y=20)$.

We find the following results. 
The accretion process is well established for small radii, $x,y <10$, indicating steady state 
accretion for $t>500$.
This is also the area from which the most energetic part of the jet is launched.
However, the accretion process has not yet established a steady state for radii larger than
$x,y = 20$ - even after 5000 dynamical time steps.
The accretion rates measured at radii $x,y = 5, 10, 15$ are about 0.02, 0.06, 0.07 (in code units), 
respectively.

The accretion rate is decreasing for smaller radii.
This confirms that part of the accreting material is diverted into the outflow 
before it reaches smaller radii.
However, we should keep in mind that there is a time delay caused by the finite advection 
velocity of material from the outer to the inner disk radii.
We should therefore be cautious when comparing the exact values for accretion and advection 
for different radii.
For typical advection velocities $ < 0.01$ at radius 15 the advection time 
scales from $r= 15$ to $r=5$ is of about 1000 dynamical time steps
which seem to lead to a missing mass\footnote{Subtracting the ejection 
rate $2\times 0.01$ integrated till $r=15$ (see Figure 11) from
the accretion rate 0.06 (see Figure 10) gives a value 0.04, larger than the 
accretion rate at $r=5$ of 0.02.}
Nevertheless, our general statement of the mass loss from accretion to ejection is correct.

The sudden growth in mass flux at distinct times, seen first at large radii and then also for smaller 
radii, we understand as triggered by the inflation and misalignment of the disk due to the 
tidal effects of the companion star. 

\begin{figure*}
\centering
\includegraphics[width=18cm]{\figurepath/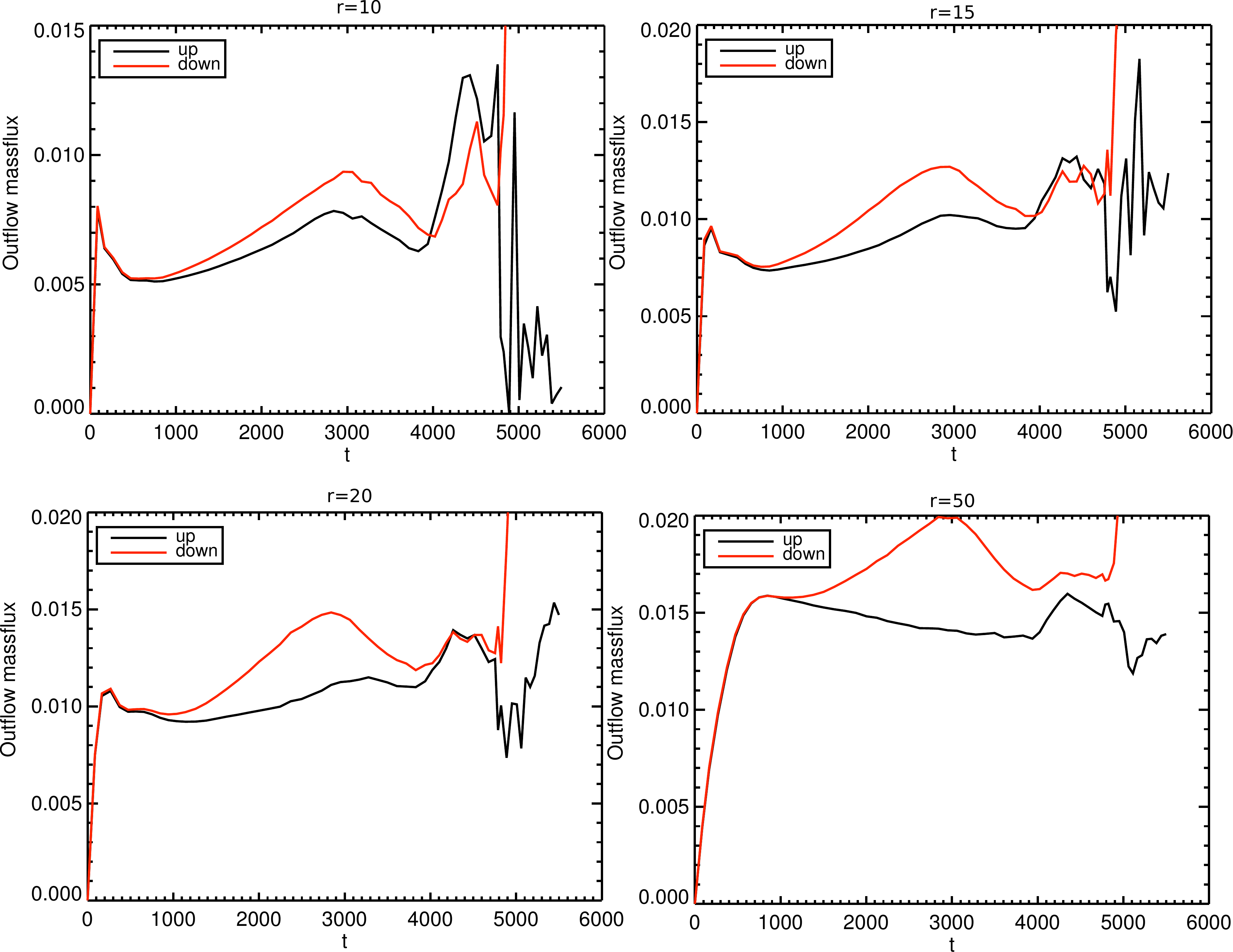}
\caption{Asymmetric outflow mass flux evolution. Shown are
the evolution of the ejected mass flux for jet (upper hemisphere, {\em red}) and
counter jet (lower hemisphere, {\em black}) for simulation {\em bcase2}.
The number values were integrated for different control volumes extending to $x,y = 10, 15, 20, 50$ 
and $z = 3H$ where $H=H(x,y=10), ..., H(x,y=50)$ is the thermal scale height of the disk the
corresponding outer radius, respectively.
 }
\label{fctbinary2_outflowrate}
\end{figure*}

In addition, we measure the outflow mass flux, $\rho v_{\rm z}$, along the $z$ axis.
Figure \ref{fctbinary2_outflowrate} shows the evolution of the outflow mass flux for 
the two hemispheres, integrated in different control volumes as specified above.
We see that the respective outflow mass fluxes into the upper and lower hemisphere are different
and that this difference grows in time.

The formation of the warp highly affects the ejection rate from the upper hemisphere.
We observe that a peak {\snr appears} in the outflow mass flux integration from the upper hemisphere.
Furthermore, this peak could be a signature of the disk misalignment compared to the initial mid-plane of the disk.

Interestingly, the peak becomes larger and larger for larger radii.
We may estimate the time scale for initiating the warp in the disk.
In our run {\em bcase2}, we recognize that the warp starts building up around $t= 1000$.
Comparing this value with run {\em bcase1} with the 
larger binary separation, we find that the warp builds up
later, around $t=2000$.

\begin{figure*}
\centering
\includegraphics[width=18cm]{\figurepath/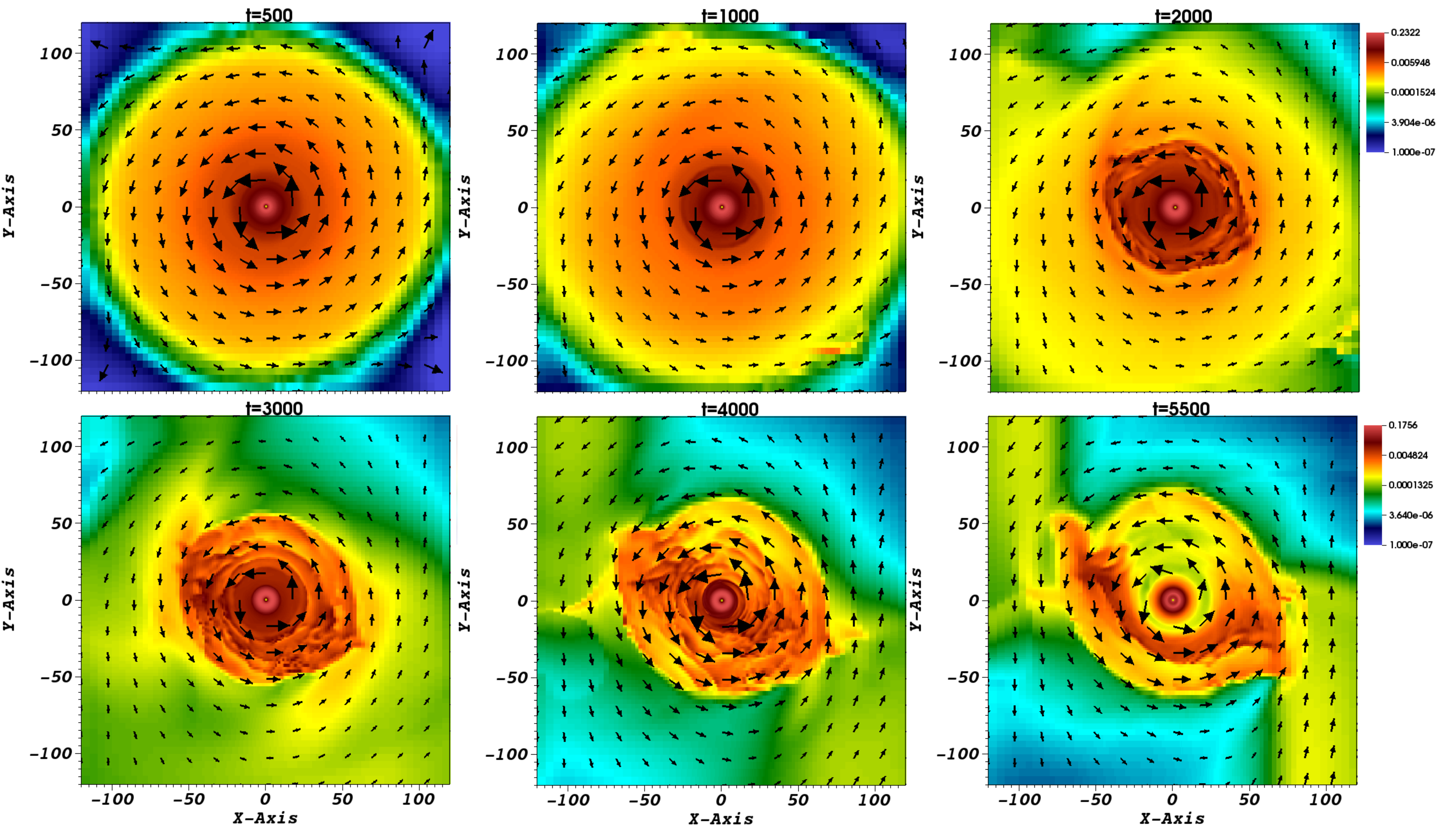}
\caption{Non-axisymmetric evolution of the disk. 
Shown are the two dimensional slices done for the mass density distribution, 
in the $x-y$ plane for bcase2 at different dynamical times.
The arrows indicate to the velocity vectors of the disk.}
\label{bcase2_rho_xy}
\end{figure*}

Figure \ref{bcase2_rho_xy} displays 2D slices of the mass density in
the $x-y$ plane for the case {\em bcase2} at different times.
Comparing Figure \ref{bcase2_rho_xy} for the simulation applying the 3D gravitational potential to
Figure \ref{fig:fct9_rho_xy} for the axisymmetric 3D setup (the test case), we observe the 
following differences.

\begin{itemize}
\item[(1)]{After dynamical time 1000, we observe that the axial symmetry is broken.
The asymmetric pattern is growing and finally leads to a structure that looks like spiral arms.}

\item[(2)]{Contrary to the simulation with the axisymmetric setup, we do not observe the rectangular pattern
in the outer disk structure at the late evolutionary stages.
Now, the circular distribution of the mass density turns into a pattern that has an elliptical shape.}

\item[(3)]{While for the test run, the area where we observe fluctuations in the disk structure
was confined to the region $40<r<60$, for the case with the very small binary separation, we see that 
the area of perturbations further extends to smaller disk radii. 
The degree of the perturbations is somewhat larger.}
\end{itemize}

\begin{figure*}
\centering
\includegraphics[width=18cm]{\figurepath/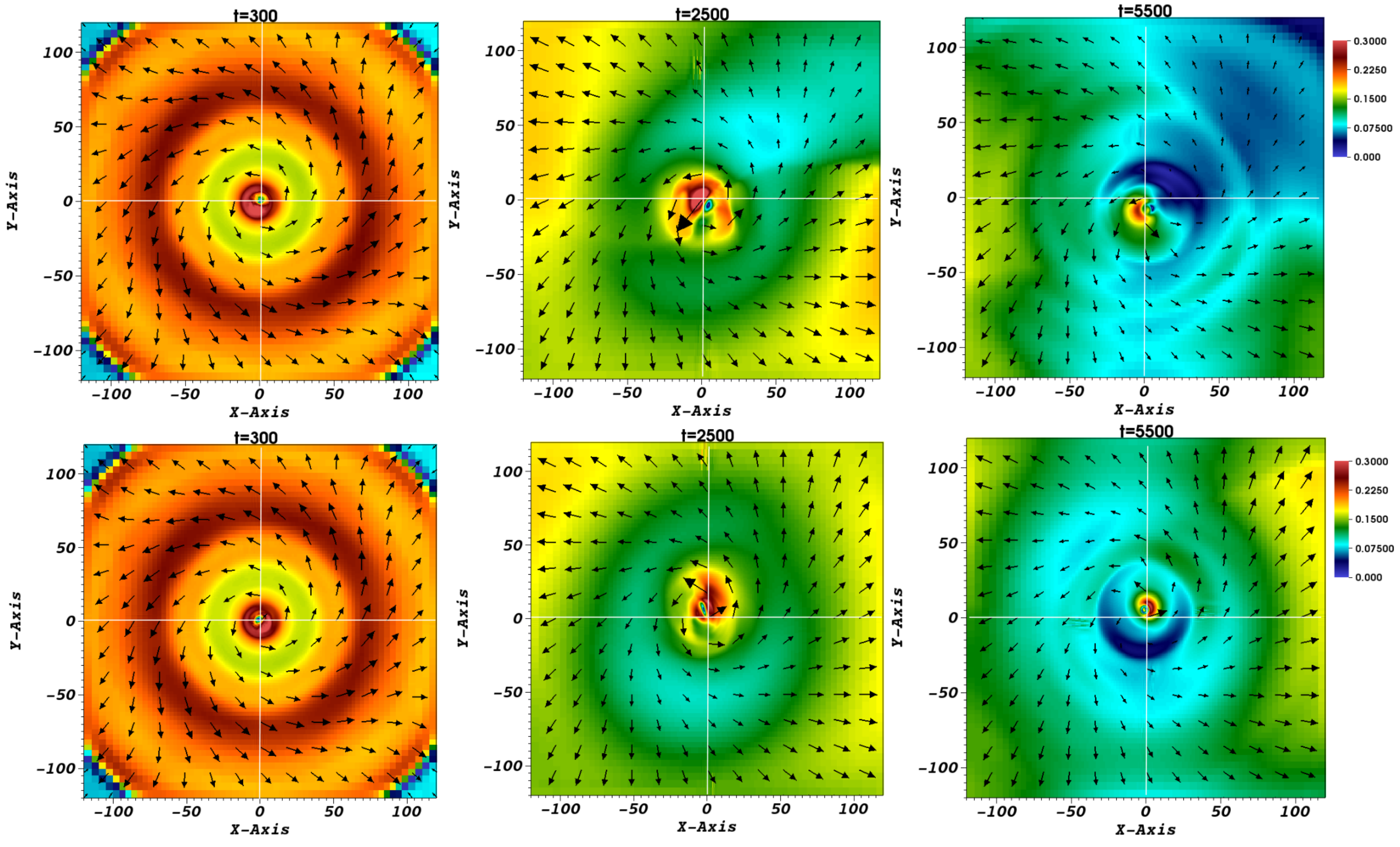}
\caption{Indication of jet precession.
Shown is the cross section of the projected velocity (color-coded) in the $x$-$y$ plane at $z=140$ 
(top, jet) and at $z=-140$ (bottom, counter jet) for the simulation {\em bcase2} at different times.
The white lines indicate the grid center, thus the initial disk/jet rotational axis.
The arrows indicate the velocity vector field.}
\label{precession_z140}
\end{figure*}

We have discussed above three 3D effects that we observed in simulation {\em bcase2} considering 
a binary star-disk-jet system with a small separation. 
That are (i) the jet inclination outside the Roche, (ii) the disk 
warping, and (iii) the indication of disk precession.
A further 3D feature is (iv) a spiral arm that develops in both the jet and the counter jet.
We observe that for jet (positive $z$) the material of the spiral arm is somewhat denser.
This is in principle understandable.
Due to our initial setup with the secondary being located above the mid-plane, the jet is more exposed 
to the gravity and torques generated by that star, and is, thus, more responding to 3D effects.

Another effect we may expect is {\em jet precession} due to the precession of the jet launching disk.
In order to find any signature of disk/jet precession, we have run simulation {\em bcase2} for about 
900 inner disk rotations - despite the mass loss of the disk involved.

Figure \ref{precession_z140} shows $x$-$y$ slices of the jet velocity 
taken at $z=140$ (top) and $z=-140$ (bottom) for the time steps $t=300, 2500, 5500$.
The $x$ and $y$ axes are indicated by the white lines.
Thus the initial outflow axis is located at the origin of the $x$-$y$ plane.
Since jet precession should be resolved easier for large distances along the jet, we focus on the 
evolution of the  jet velocity at large height, $z=140$.
When the system evolves in time, the jet axis (the blue colored region) moves away from
the initial jet axis.
At time 5000, the offset is about $10\,R_{\rm i}$ corresponding to a $\simeq 4.0$ degree opening angle
of the precession axis.
Essentially, is the jet axis {\em moves along an arc} of 4 degrees length in the $x-y$ plane.  
We argue that due to this 2D motion, the offset of the jet axis cannot just be a projection effect of the 
jet axis affected by inclination.
Instead, it suggests the initiation of the precession of the jet axis across the $x-y$ plane.
In the mass distribution across the asymptotic jets (not shown), we also find a deviation from axial 
symmetry.
Considering also the fact that also the disk alignment has changed with respect to the initial disk 
mid-plane, 
{\em we interpret the offset of the jet axis as strong indication for the onset of jet precession - caused 
by the precession of the jet launching disk.}

We may estimate the time scale of jet precession. 
For the setup with small binary separation {\em bcase2}, to reach about 4 degree opening angle,
we run the simulation for 5000 dynamical, corresponding 24 years.
However, theoretical estimates (see above) suggest for the full precession of the jet about 20 orbital
time scales corresponding to 1400 years in our case.
Thus, much longer simulations are required to fully disentangle jet or disk precession effects.
With our specific parameter setup, we might, however, have detected initial signatures of disk/jet 
precession in our non-axisymmetric 3D model setup.

Concerning the bipolar symmetry, the jet velocity maps show that the counter jet is more collimated than 
the jet.
The counter jet maintains the axial symmetry rather well, while the jet is deviating from the axisymmetric 
structure evolving into cross section of rather elliptical shape.
More over, the counter jet does not exhibit a remarkable offset between the rotation axis and 
the grid center.
As mentioned in the previous section when we discussed the jet inclination outside the Roche lobe, this hints 
on weaker tidal affects on the counter jet, while the jet itself is more disturbed by the gravity
of the secondary. 
This is an effect on top of the disk / jet precession.

\section{Conclusions}
We have presented results of numerical simulations studying the three-dimensional jet launching from 
a diffusive accretion disk threaded by a large scale magnetic field.
We have hereby extended our previous axisymmetric model setup to fully 3D.
Essential modifications had to be made concerning the inner boundary conditions that have to work as an 
internal accretion boundary (a "cylindrical sink" for mass and angular momentum), and also the outer disk 
rotation close to the outer boundary of the rectangular computational grid. 
In order to establish a proper rotation of the inner jet launching disk we have prescribed a sub-Keplerian 
rotation along the ghost cells within the internal boundary.
Further, a non-rotating corona is surrounding a disk of finite radius.
We have obtained the following results.

(1) Our reference run {\em scase2}, considering a single star (thus an axisymmetric gravitational potential),  
was run for about 600 rotations (about 4000 dynamical time steps) at the inner disk radius. 
We find bipolar jets that are launched from the inner part of the disk and are accelerated to 
super Alfve\'nic and super fast speed. 
The overall large-scale outflow structure shows a well-kept right-left (rotational) symmetry and also a good 
bipolar (hemispheric) symmetry of the outflow, approving the quality of our 3D model setup.
Due to the different prescription for the magnetic diffusivity and also a higher 
numerical diffusivity (given the lower resolution in 3D), we obtain higher accretion rates.
The rough number of the mass fluxes are, however, comparable with those obtained in the
axisymmetric setup \citep{2012ApJ...757...65S}.
The accretion-ejection mass flux ratio is somewhat higher than for the 2D simulations.
Clearly, the exact numbers depend on the further parameter choice such as magnetic field strength
and magnetic diffusivity.

(2) As a next step, we have implemented the gravitational potential of a binary system in our 3D 
reference model and have run simulations with a variety of parameter choices.
In this setup, we were able to observe disk warping and consecutive jet inclination and the initial signatures 
of disk and jet precession.
Since precession effects typically establish on longer time scales than we can run our simulations,
we have setup simulation {\em bcase2} that applies a smaller binary separation of $200\,R_{\rm i}$
together with an initial orbital inclination in order to amplify the tidal effects.
Due to the limited disk mass (the inner Lagrange $L_1$ point is located in the computational domain
for run {\em bcase2})
the running time of the simulation was limited to 900 inner disk rotations 
(equivalent to 5000 dynamical time steps).
Nevertheless, we were able to disentangle several non-axisymmetric tidal effects that are expected from
a 3D model setup considering a binary system.

(3) The structure of the accretion disk is affected strongly by the tidal forces in the binary system. 
The part of the disk close to the inner Lagrange point $L_1$ slowly expands beyond the Lagrange point. 
This can be understood as initialization of a Roche lobe overflow, a feature that is well established 
in simulation {\em bcase2} with a small binary separation in particular.
Moreover, the alignment of the accretion disk with respect to the initial mid-plane changes.
The combination of these effects build up a {"}bump{"} in the disk region that is closer to the 
companion star. 
The formation of the bump is the first signature of disk warping.
In fact the bump that forms will later be part of the disk warp.

(4) The time evolution of the mass fluxes calculated for different radii shows that the warp first 
forms at the outer disk and then moves inwards. 
The warps appear as sudden peaks in the accretion rate, visible consecutively at different 
radii - first at large radii and with some time lag also at smaller radii. 
In our simulation {\em bcase2} with a small binary separation, we recognize that the disk warp builds
up after about $\simeq 160$ inner disk rotations (corresponding to about 5 years for protostars).
For the simulations with larger binary separation {\rm bcase1}  (and thus larger orbital periods),
the tidal effects are weaker and a warp is formed later, only at about 1200 dynamical time steps
($\simeq 190$ inner disk rotations).

(5) A further non-axisymmetric effect is jet inclination - the deflection of the jet motion from the 
initially axial motion along the $z$-axis.
Jet inclination results from the global force balance affecting the jet material when it has left the 
Roche lobe of the primary.
Consequently, in the case when the secondary is located in the upper hemisphere, stronger jet inclination 
is seen for the jet propagating into this hemisphere.

(6) The structure and the alignment of the accretion disk changes.  
The disk becomes increasingly more misaligned with respect to the initial disk mid-plane and tends 
to align with the orbital plane of the binary.
Considering the jet velocity far away from the launching area, we find that the jet rotation axis 
moves along an arc of 4 degree in the $x-y$ plane.
This holds as well for the jet density.
The effect appears only if the disk initial mid-plane and the orbital plane are misaligned.
Altogether we interpret these effects as strong indication for the onset of {\em disk precession}.
Due to the high computational costs, we were are not yet able to observe a full precession cycle 
during the run time of our simulations.

(7) The most intriguing non-axisymmetric effect we observe in our simulations is the onset
of {\em jet precession} as a consequence of the disk precession.
Precession establishes on much larger time scales than we can run our simulations - on orbital time 
scales, some 100 times longer than our simulation runs.
However, for our model setup of a close binary {\em bcase2} we find clear indication
of jet precession in its initial stages.
Considering slices of the jet velocity across the jet and counter jet,
we observe that the jet rotation axis moves away from its initial alignment along the vertical axis.
If our interpretation of an initial disk and jet precession is correct, we may thus quantify the 
precession cone of the jet axis by an opening angle of about 4 degree - measured after 5000 
dynamical times (corresponding to 24 years for YSOs, or 7 years for AGNs).
In order to follow jet precession fully, a much longer simulation would be necessary.
This is currently impossible due to the limited mass reservoir of the accretions disk.

\smallskip
In summary, we have shown the non-axisymmetric evolution of the disk-jet launching process
applying magneto hydrodynamic simulations.
In particular, by considering jet launching in the Roche potential of a binary system we have demonstrated 
a number of non-axisymmetric effects in the disk-jet system evolution, in particular disk warping and 
jet inclination.
Simulations treating a jet-launching disk misaligned with the binary orbital plane were able to trace 
the onset of disk precession - instantly also resulting in a jet precession.
Our simulations numerically confirm that tidal forces is significant for generating jets that are inclined
or precessing, and accretion disks that are warping.

\acknowledgements
We thank Andrea Mignone and the PLUTO team for the possibility to use their code.
We thank Rachid Ouyed for valuable comments.
Our simulations were performed on the THEO cluster of the Max Planck Institute for Astronomy
and the HYDRA cluster of the Max Planck Society.
This work was partly financed by the SFB 881 of the German science foundation DFG
and by the institute for research in fundamental Sciences (IPM) .
We thank an unknown referee for suggestions that improved the presentation of the paper.

\appendix

\section{A. Specific boundary conditions}
\label{sec:BC}
It is essential for the simulations to define a smooth and axisymmetric boundary condition along the 
{``}radial{''} boundary of the sink.
There are two major points to be considered.
That is, firstly, the possibility that the axisymmetric evolution of the inner disk and jet may be artificially 
disturbed by the rectangular grid.
Secondly, we cannot simply imply the standard PLUTO outflow condition that copies the grid-internal values 
onto the ghost cells of the internal boundary (the sink). 
There is just no clear way how the grid-internal values can be copied consistently, as certain 
ghost cell would receive a copy from different grid-internal cells.
This difficulty is most essential for the velocity vector describing rotation and accretion.

As mentioned above, for the inner boundary
we make use of the {\em internal boundary} option of PLUTO.
We prescribe rectangular structure of ghost cells within the active domain
and apply user defined boundary values that allow to absorb disk material
and angular momentum that is advected to the inner disk radius and that ensure
an axisymmetric rotation pattern in the innermost disk area
(see Figure \ref{fig:innerBC}, left).

The boundary condition at the top and the bottom of the sink prescribes the initial
value for the gas pressure and a density of 115\% the initial local density, in order
to avoid the evacuation of the 
region close to the rotation axis.
Effectively, this boundary condition replenishes some mass into the domain and therefore avoids low densities close to the rotation axis.
For the velocity, we assign an injection into the domain of low velocity above and below the sink,
$v_z = \pm 0.01$.
As a consequence, matter is injected close to the rotational axis with a small background velocity
avoiding infall of matter into the sink.
The injected low density material accumulates to a the mass flux out of the sink about 1000 times less 
than main jet launched by the disk. 

\subsection{Inner boundary}
Adjacent to the inner radius $r_{\rm i}$ of the disk, i.e the cylindrical sink,
we adopt a cylindrical shell of four ghost cells thickness.

\begin{figure}
\centering
\includegraphics[height=6.5cm]{\figurepath/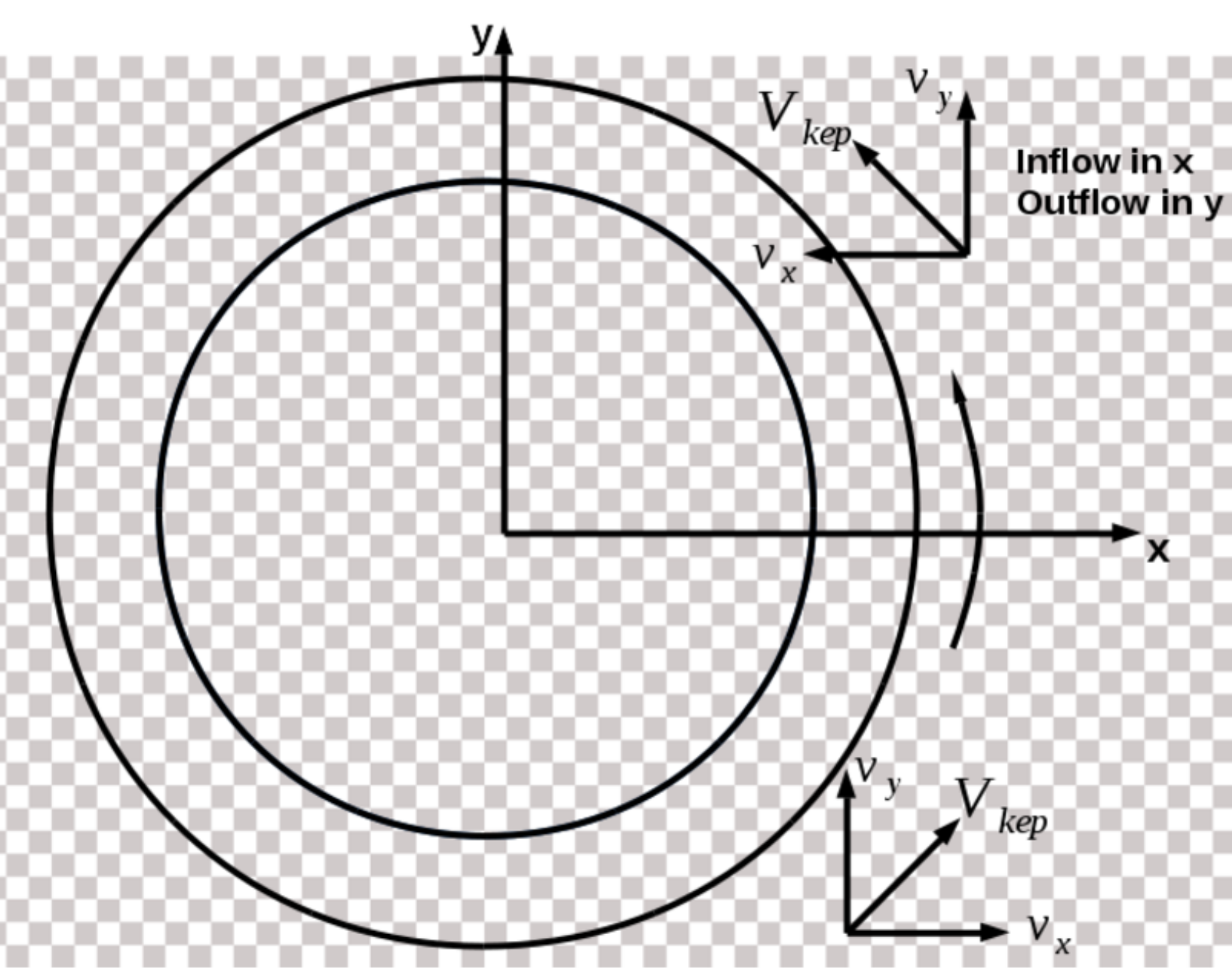}
\includegraphics[height=6cm]{\figurepath/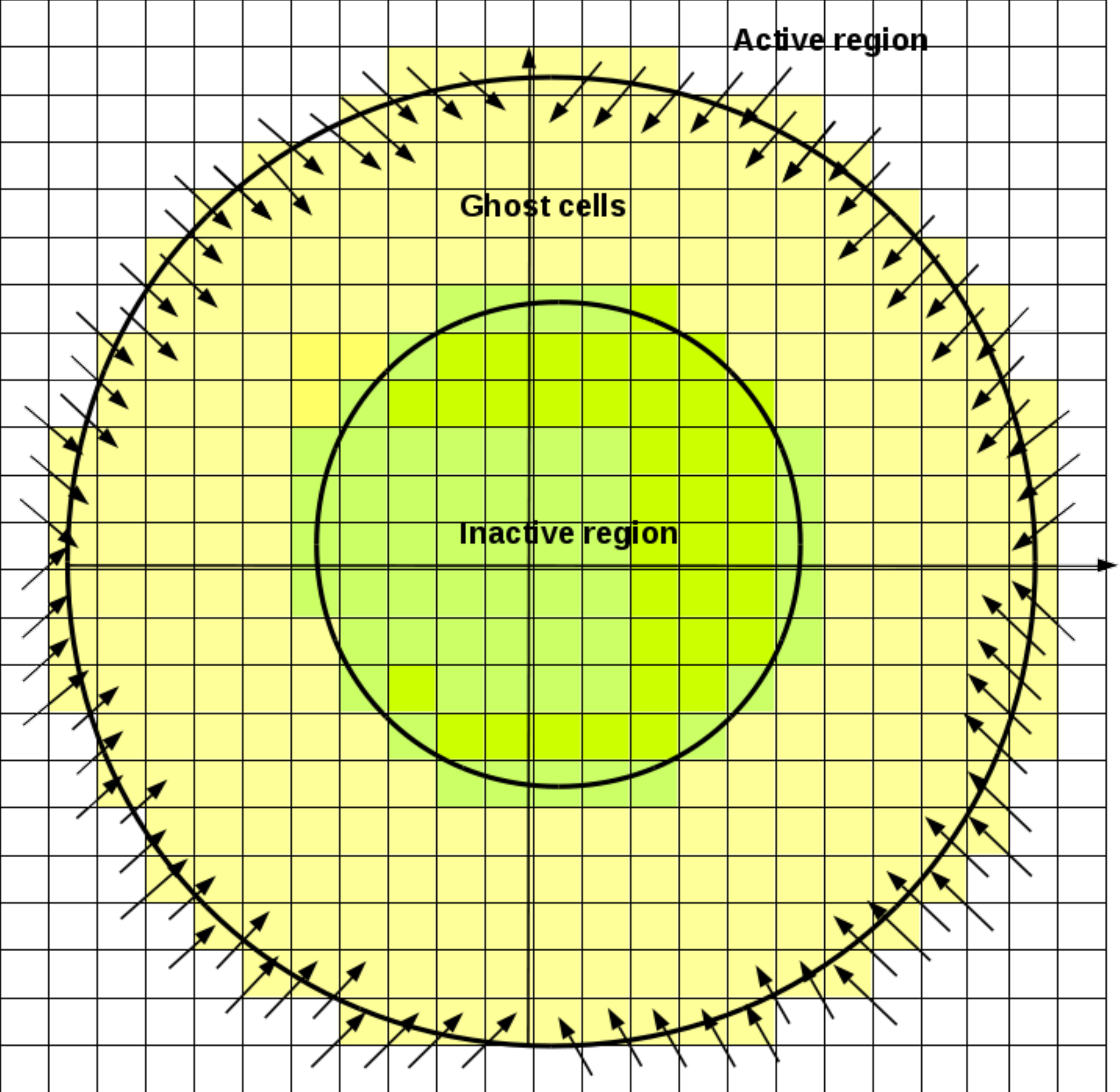}
\caption{
{\em (Left)}  The area of the inner disk boundary. 
The boundary is defined as an {\em internal} boundary condition, with ghost zones between the inner disk boundary 
(outer ring) at radius $r=1$ and the ghost area (limited by the inner ring). 
The boundary conditions are defined on grid cells between this rings (actually cylinders of 
height $0.8$).
The difficulty to copy boundary values in radial direction consistently from cells 
outside the ring to cells inside the ring is clearly visible.
The velocity vectors indicate the difficulty to disentangle accretion velocity from orbital motion.
{\em (Right)} Sketch of the way we copy the densities and pressure and vertical velocity $v_z$.
The copying is done diagonally from the active domain to the ghost cells.
}
\label{fig:innerBC}
\end{figure}

These ghost cells are used to absorb mass flux and angular momentum from the accreting
material.
In order to do so, at each angular position along the shell, four diagonal cells are used 
to copy certain hydrodynamical variables from the active domain into the ghost cell.
In particular, we copy the values for density, gas pressure and the vertical velocity $v_{\rm z}$,
from outside the boundary into the ghost cells.
The copying is done in radial direction, from the cell $(i \pm 1,j \pm 1)$ into the ghost cell 
$(i, j)$, and similarly for ghost cells further in (see Figure \ref{fig:innerBC}, right).

For the other velocity components, $v_{\rm x}$ and $v_{\rm y}$, such a procedure of copying or extrapolating 
internal values onto the ghost cells is not feasible, as the angular momentum conservation is easily violated.
Figure \ref{fig:innerBC} (left) visualizes these difficulties.
The figure shows a $x$-$y$ slice of the grid in the mid-plane of the domain and the corresponding vectors of 
the rotational velocity along the boundary.

In practice, in Cartesian coordinates the toroidal velocity component is described by the $x$ and $y$ components
of the velocity vector, $v_{\phi} = -sin{\phi} v_x + cos{\phi} v_y$, where $\phi$ indicates to the velocity direction 
in the $x$-$y$ plane, $\phi=\arctan{v_y/v_x}$.
The difficulty is that these velocity components at the same time also describe a {\em radial motion} 
(advection, disk accretion) that could not easily be disentangled from orbital motion.
We therefore decided to develop another approach, by that we essentially
{\em prescribe} a disk rotation within the ghost cells along the internal boundary. 
The orbital motion in the ghost cells is chosen to be slightly lower than the velocity of the disk.
This corresponds to a rotation with lower angular momentum and therefore allows to absorb the
material that has been advected by the disk (see Equation \ref{vphi_profile}).

Concerning the boundary conditions for the magnetic field, we have first tried to extend our 2D setup 
to 3D also for magnetic field boundary condition along the inner boundary (i.e. specific prescription
for the advection of vertical magnetic flux and electric current conservation across the sink boundary, 
see Sheikhnezami et al. 2011). 
We found, however, that such an approach would introduce a unreasonable amount of $\vec{B}$ if 
we apply a similar copying procedure as for the velocity.
We have therefore decided to follow a different approach.
That is that we evolve the magnetic field in the ghost cells of the internal boundary as in 
the active domain. The code treats the internal ghost cells like active cells. However, we do
not overwrite the magnetic field vectors with a boundary value, as we do for the velocity.
This is insofar a kinematic approach for the magnetic field boundary condition.
Since we do not evolve the hydrodynamic state within the internal boundary in time, the
hydrodynamic evolution is decoupled from the field evolution.

The sink / internal boundary still allows for the conservation of magnetic flux and electric 
current considering constrained transport, but has no other hydrodynamic influence than 
acting as a sink of matter and (angular) momentum.

\subsection{Outer boundary}
A similar problem concerning the rotational velocity occurs at the outer boundary of the 
(rectangular) grid.
Along the same side of the boundary, the orbiting disk material is supposed to flow out across 
one part of the boundary and then to flow in again across another part of the boundary 
(considering either the $x$- or the $y$-component of the velocity vector, see Figure \ref{fig:innerBC}, 
left).

This problem is well known and has been discussed by other authors who studied the 3D structure 
of the outflow formation \citep{2003ApJ...582..292O, 2013MNRAS.429.2482P}.
These authors discuss the problem of the dual boundary conditions for the velocity components,
i.e. inflow / outflow, in a Cartesian grid, and finally apply the condition of vanishing toroidal 
velocity in the outermost and in the innermost grid area, thereby implementing a {\em finite} disk 
size.

This strategy works well for us for the outer boundary \footnote{It does not - unlike 
for the cited literature - work for the inner part of the disk.  
The reason is that we consider the {\em launching} problem and have to take care of the advection of 
material across the inner boundary.
This inner boundary is particularly essential for the accretion process in the disk.}, where
we initially prescribe a static disk structure (see Equation \ref{vphi_profile}).
However, we note that as time evolves, the disk rotation spreads out also to larger radii, and some
disk mass will be lost on the long time scales.

For the outer boundaries of the computational domain, standard outflow conditions are applied as
prescribed by PLUTO.
Such zero-gradient outflow conditions may however lead to an artificial re-collimation of the flow as
suggested by \citet{1999ApJ...516..221U}.
We have therefore also applied modified outflow conditions that have been suggested by \citet{2010ApJ...709.1100P} 
and have been used previously \citep{2012ApJ...757...65S, 2013ApJ...774...12F}. 
However, given the relatively short time scale of our 3D simulations, we did not observe a difference in 
the collimation degree and therefore decided, for simplicity, to apply the standard outflow conditions provided
by PLUTO in the present paper.

\section{B. Observed jet sources with binary signature}
The typical configuration of a jet-launching star - a magnetized star-disk system - is also
found in {\em evolved} binary systems. 
These sources are known as Low-Mass or High-Mass X-ray Binaries, Cataclysmic Variables, or 
micro-quasars 
\citep{1984ARA&A..22..507M, 2006ApJ...651..416F, 2011IAUS..275..311T}.
However, only very few jet sources are known for these cases and for some classes, such as Cataclysmic
Variables, the indication for jets is still controversial \citep{2005ASPC..330..117L, 2011MNRAS.418L.129K}.

In Table \ref{tbl:binary star-observation}, the physical parameters of some observed binary systems 
are collected.
For comparison, the parameters for some close binary systems as e.g. High Mass X-ray Binaries and Cataclysmic 
Variables binaries are also shown.

\begin{table*}[h]
\caption{Examples of binary stars. Displayed are selected observed physical parameters as
the mass of the primary $M_{\rm p}$,
the mass of the secondary $M_{\rm s}$, 
the binary separation between $D_{\rm s}$, 
and the orbital period $T_{\rm orb}$.
For comparison, 
parameters for some close binary systems such as High-Mass X-ray Binaries and Cataclysmic 
Variables binaries are also shown.
}
\begin{center}
 \begin{tabular}{lccccc}
\hline
\hline
\noalign{\smallskip}
 Object         & $M_{\rm p}$               & $M_{\rm s}$             & $D_{\rm s}$     &  $T_{\rm orb}$  \\
 \noalign{\smallskip}
 \hline
 \noalign{\smallskip}
 Young Stars (YSOs)\\
 \noalign{\smallskip}
  \hline
\noalign{\smallskip}
 Alpha Centauri & 1.1 $ M\odot $      & 0.907 $M\odot$    &  11.4-36.0 AU  & $79.91\pm 0.011$ yrs \\ 
 61 Cygni A & 0.7 $ M\odot $    & 0.63 $ M\odot $   &  44-124 AU     & $ 678 \pm 34 $ yrs \\
 RW Aur A & 1.3 – 1.4  $ M\odot $ &  0.7 – 0.9$ M\odot$ &  170 AU        &  $ > 700$ yrs    \\
 T Tau    & $M_{\rm p}+ M_{\rm s}=5.3 M\odot $  &                   &   7-15 AU&   38.8 yrs    \\
 HK Tau    & ?  &       ?            &   386AU      &      \\
\noalign{\smallskip}
  \hline
\noalign{\smallskip}
Cataclysmic Variables (CVs)\\
\noalign{\smallskip}
  \hline
\noalign{\smallskip}
BV Cen  & $ 0.83 M_\odot$ &  $ 0.90 M_\odot$ &        &   0.611 days    \\
Hu Aqr  & $ 0.95 M_\odot$ &  $ 0.15 M_\odot$ &        &   0.08682 days    \\
\noalign{\smallskip}
  \hline
\noalign{\smallskip}
 High Mass X-ray Binaries (HMXBs)\\
\noalign{\smallskip}
  \hline
\noalign{\smallskip}
Vela X-1  & Ns           & Super giant   &   &  8,96 days\\
Cyg X-1   & BH Candidate & Super giant                     &  & 5.60 days \\
SS 433    & BH or Ns     & Super giant                       &  & 13.1 days \\
\noalign{\smallskip}
  \hline
\noalign{\smallskip}

 \end{tabular}
\end{center}
\label{tbl:binary star-observation}
\end{table*}

\section{C. Disk evolution of binary star-disk-jet system {\em bcase1}}
\begin{figure*}
\centering
\includegraphics[width=18cm]{\figurepath/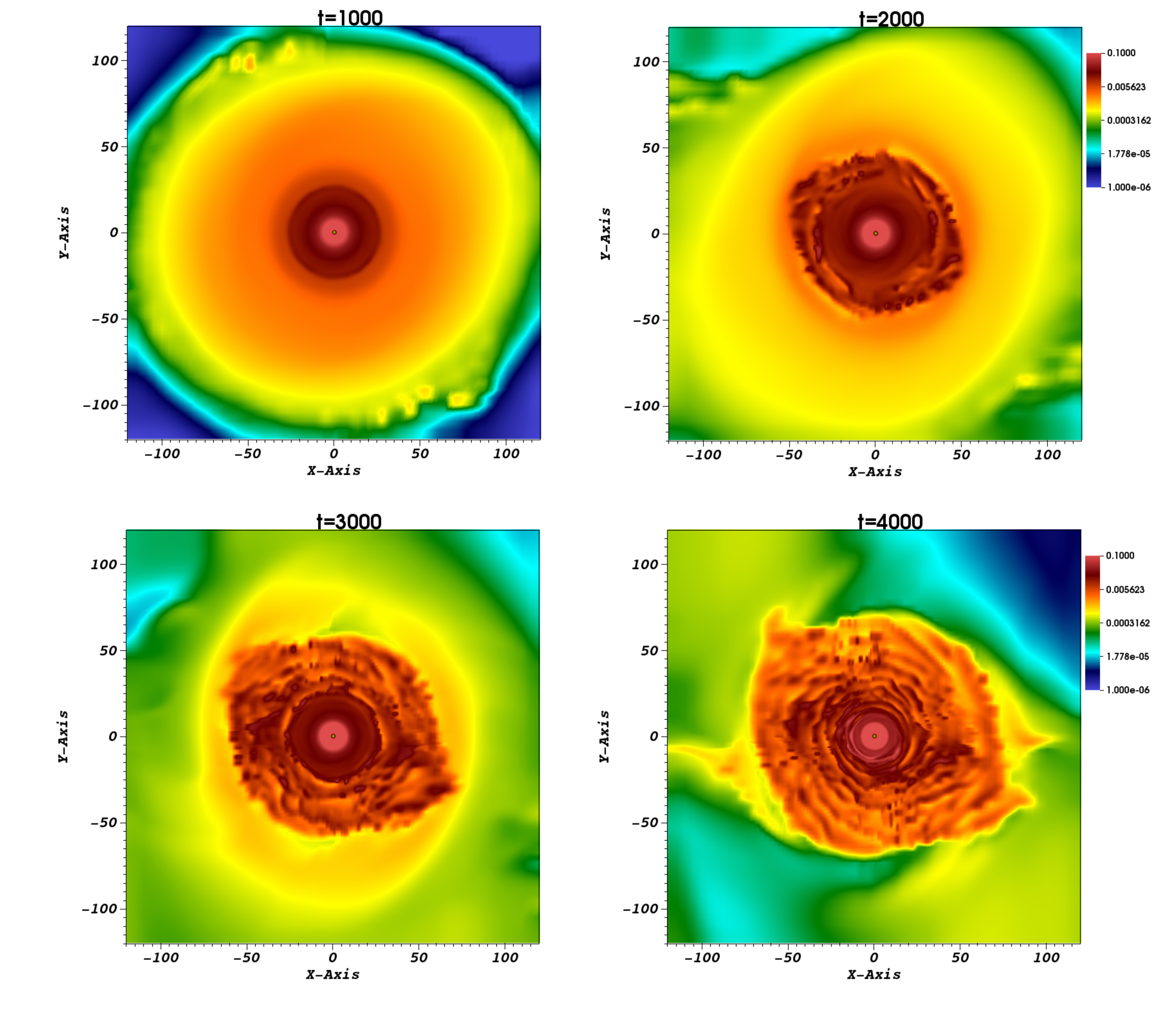}
\caption{Shown are the two dimensional slices done for the mass density distribution, 
in the $x-y$ plane for bcase1 at different dynamical times.}
\label{bcase1equa}
\end{figure*}

For comparison, we show in Figure \ref{bcase1equa} 2D slices of the density distribution in
the $x$-$y$ plane (initial disk mid-plane) for simulation run {\em bcase1} at different times.

Compared to the evolution of {\em bcase2}, as shown in Figure \ref{bcase2_rho_xy},
it looks that the density perturbations are more prominent in {\em bcase1}.
As we discussed above, one possible cause for these fluctuations may be the development of the 
magneto-rotational instability in these outer parts of the disk.
Since the magnetic diffusivity plays a major role for the onset of the MRI, it is worth to emphasize, 
that the magnetic diffusivity prescription is different in both cases.
In {\em bcase1} the diffusivity is confined to the disk area, while in {\em bcase2} a background 
diffusivity was defined for the whole grid.
However, it seems that other physical effects such as magnetic field strength or even tidal forces 
should also affect the existence of the fluctuations.

\bibliographystyle{apj}
\bibliography{mypaper}

\end{document}